\documentclass[preprint,8pt,3p,fleqn]{elsarticle}
\usepackage{amsmath}
\usepackage[T1]{fontenc}
\usepackage{graphics}
\usepackage{inputenc}
\usepackage{graphicx}
\usepackage{epsfig}
\usepackage{amsmath,amssymb}
\usepackage{esint}
\usepackage{float}   
\usepackage{caption}
\usepackage[usenames, x11names]{xcolor}
\usepackage{color, colortbl} 
\usepackage{multirow}
\usepackage{relsize}
\usepackage[colorlinks=true,linkcolor=black,filecolor=blue,citecolor=red,bookmarksnumbered=true]{hyperref}
\usepackage{setspace}
\usepackage{booktabs}
\usepackage{tabu}
\usepackage{makecell}
\usepackage{adjustbox} 
\usepackage{subcaption}
\usepackage{enumitem}
\biboptions{sort&compress}
\usepackage{lineno}
\usepackage{ifthen}
\usepackage{bm}
\usepackage{ulem}
\newcommand{\tensor}[1]{{\smash{\uuline{#1}}}}
\usepackage{etoolbox,xspace}

\newcommand\vecu{\mathbf{u}}
\newcommand\vecv{\mathbf{v}}
\newcommand\vecw{\mathbf{w}}
\newcommand\vecU{\mathbf{U}}
\newcommand\vecf{\mathbf{f}}
\newcommand\vecF{\mathbf{F}}

\newcommand\tk{t^{_{(k)}}}
\newcommand\tkk{t^{_{(k+1)}}}

\DeclareFontFamily{OT1}{pzc}{}
\DeclareFontShape{OT1}{pzc}{m}{it}{<-> s * [1.10] pzcmi7t}{}
\DeclareMathAlphabet{\mathpzc}{OT1}{pzc}{m}{it}
\DeclareFontEncoding{LS1}{}{}
\DeclareFontSubstitution{LS1}{stix}{m}{n}
\DeclareMathAlphabet{\mathscr}{LS1}{stixscr}{m}{n}
\SetMathAlphabet{\mathscr}{bold}{LS1}{stixscr}{b}{n}
\DeclareFontFamily{U}{mathc}{}
\DeclareFontShape{U}{mathc}{m}{it}{<->s*[1.03] mathc10}{}
\DeclareMathAlphabet{\matholdcal}{U}{mathc}{m}{it}
\newcommand\identity{\bm{\mathcal{I}}}
\newcommand\vecg{\bm{\mathscr{s}}}
\newcommand\vecG{\bm{\mathscr{S}}}
\newcommand\vecpd{\bm{\mathscr{l}}}
\newcommand\vecPD{\bm{\mathscr{L}}}
\newcommand\vecO{\bm{\mathscr{O}}}
\newcommand\K{\matholdcal{K}}
\newcommand\B{\matholdcal{b}}

\newcommand\vfrac{\varepsilon}

\newcommand\vecx{\mathbf{x}}
\newcommand\vecy{\mathbf{y}}
\newcommand\vecz{\mathbf{z}}
\newcommand\vecr{\mathbf{r}}

\newcommand\vecq{\mathbf{q}}
\newcommand\vecX{\mathbf{X}}
\newcommand\vecY{\mathbf{Y}}

\newcommand\erf{\,\mathrm{erf}}

\newcommand\hh{\matholdcal{h}}
\newcommand\Hh{\matholdcal{H}}
\newcommand\dd{\matholdcal{d}}
\newcommand\D{\matholdcal{D}}
\newcommand\ie{i.e.\xspace}
\newcommand\Conv{\circledast}

\usepackage[T3,T1]{fontenc}
\DeclareSymbolFont{tipa}{T3}{cmr}{m}{n}
\DeclareMathAccent{\invbreve}{\mathalpha}{tipa}{16}
\newcommand{\integral}[1]{%
  \ensuremath{{{#1}}}%
}

\journal{Journal}

\begin{document}
	
\begin{frontmatter}

\title{Undisturbed velocity recovery with transient and weak inertia effects in volume-filtered simulations of particle-laden flows} 

\author[label1]{Fabien Evrard}\corref{cor1}
\cortext[cor1]{Corresponding author}
\ead{fevrard@illinois.edu}
\author[label2]{Akshay Chandran}
\author[label3]{Ricardo Cortez}
\author[label2]{Berend van Wachem}

\address[label1]{Department of Aerospace Engineering, University of Illinois Urbana-Champaign, Urbana, IL 61801, United States}
\address[label2]{Chair of Mechanical Process Engineering, Otto-von-Guericke-Universit\"at Magdeburg, 39106 Magdeburg, Germany}
\address[label3]{Mathematics Department, Tulane University, New Orleans, LA 70118, United States}

\begin{abstract}
In volume-filtered Euler-Lagrange simulations of particle-laden flows, the fluid forces acting on a particle are estimated using reduced models, which rely on the knowledge of the local \textit{undisturbed flow} for that particle. Since the two-way coupling between the particle and the fluid creates a local flow perturbation, the filtered fluid velocity interpolated to the particle location must be corrected prior to estimating the fluid forces, so as to subtract the contribution of this perturbation and recover the local undisturbed flow with good accuracy. In this manuscript, we present a new model for estimating a particle's self-induced flow disturbance that accounts for its transient development and for inertial effects related to finite particle Reynolds numbers. The model also does not require the direction of the momentum feedback to align with the direction of the particle's relative velocity, allowing force contributions other than the steady drag force to be considered. It is based upon the linearization of the volume-filtered equations governing the particle's self-induced flow disturbance, such that their solution can be expressed as a linear combination of regularized transient Stokeslet contributions. Tested on a range of numerical cases, the model is shown to consistently estimate the particle's self-induced flow disturbance with high accuracy both in steady and highly transient flow environments, as well as for finite particle Reynolds numbers.\\

\noindent © 2024. This manuscript version is made available under the CC-BY-NC-ND 4.0 license.\\
\noindent \url{http://creativecommons.org/licenses/by-nc-nd/4.0}
\end{abstract}
\end{frontmatter}

\section{Introduction}

Simulating particle-laden flows of practical interest often calls for the volume-filtering of the equations governing the fluid flow, so as to minimize the number of numerical degrees of freedom needed to reach an adequate level of accuracy. Such filtering was formalized several decades ago \cite{Anderson1967,Ishii2006}, and is commonly referred to as volume-filtered Euler-Lagrange (VF-EL) or simply Euler-Lagrange (EL) modeling, when individual particles are kept track of~\cite{Balachandar2010,Pepiot2012,Capecelatro2013,Balachandar2019}. In VF-EL simulations, the sub-filter scales of the flow, which typically include the details of the flow around individual particles, are not resolved and require modeling. Among the terms requiring closure is the momentum exchange between the Lagrangian particles and the Eulerian fluid. A two-way coupling indeed exists between the particles and the fluid, in which the fluid exerts a force and torque on the particles and the particles exert the opposite force and torque on the fluid \cite{Elghobashi1994}. Since the details of the flow around individual particles is not known in VF-EL simulations, fluid forces and torques must be estimated from the knowledge of the filtered flow using reduced models.\medskip

In the limit of vanishing Reynolds number and infinitely large flow domain, the motion of a particle immersed in a non-uniform flow is governed by the integro-differential Maxey-Riley-Gatignol~(MRG) equation~\cite{Maxey1983,Gatignol1983}. The MRG equation is classically extended to finite Reynolds number regimes through the introduction of empirical correction factors~\cite{Schiller1933,Clift1971,Clift2013} and lift force contributions~\cite{Saffman1965}. In the MRG or extended-MRG equation, all force contributions are expressed in terms of the \textit{undisturbed flow}, which is the conceptual modification of the particle-laden flow in which the particle under consideration is removed and replaced by fluid. The filtered flow, solution to the volume-filtered governing equations, is evidently unlike the undisturbed flow since the momentum fed back to the fluid by the particle creates a local \textit{self-induced flow perturbation} that is not present in the undisturbed flow. The magnitude of this self-induced velocity disturbance has been known to increase with the ratio $d_n/\max(\ell_\mathcal{K},\ell_{\Delta x})$, where $d_n$ is the particle diameter, $\ell_\mathcal{K}$ is the length-scale of the filtering kernel, if one is explicitly used, and $\ell_{\Delta x}$ is the filtering length-scale related to the discretization~\cite{Boivin1998,Balachandar2009,Evrard2021}. In traditional VF-EL approaches, such as the Particle-Source-In-Cell (PSI-CELL) method of \citet{Crowe1977}, $\ell_\mathcal{K} = \ell_{\Delta x}$ since the computational mesh acts as the filter. Only recently has the research community started to consider filter length-scales that are larger than that of the computational mesh, $\smash{\ell_\mathcal{K} > \ell_{\Delta x}}$, in an effort to mitigate mesh dependency~\cite{Pepiot2012,Capecelatro2013,Poustis2019,Evrard2019}.\medskip

Over the past years, the amount of scientific literature on the recovery of the undisturbed flow velocity from the knowledge of the filtered flow has significantly increased. The models that have been proposed commonly provide means to estimate the velocity disturbance induced by a particle as
\begin{equation} 
  \vecu^\prime = \text{Model}\left(\bar{\vecu}, \mathrm{Re}_\ell, \ell_\mathcal{K}, \ell_{\Delta x}\right) \, ,
\end{equation}
where $\smash{\bar{\vecu}}$ is the filtered velocity solution to the volume-filtered governing equations and $\smash{\mathrm{Re}_\ell}$ some definition of the Reynolds number, such that the undisturbed velocity associated with the particle can be recovered~as
\begin{equation}
  \tilde{\vecu} = \bar{\vecu} - \vecu^\prime \, .
\end{equation}
Early attempts at such modeling are based on the steady Stokes flow solution of the particle's self-induced flow disturbance \cite{Pan1996,Maxey2001,Lomholt2003}. More recently, \citet{Gualtieri2015} propose a model based on the solution of the unsteady Stokes equations with a time regularization treating singularities. Their approach is extended to wall-bounded flows in~\cite{Battista2019}. \citet{Horwitz2016,Horwitz2018} propose a model approximating the particle's self-induced velocity disturbance as a truncated power series expansion, whose coefficients are empirically fitted to data. \citet{Ireland2017} explicitly filter the analytical solution of the Stokes flow around a sphere to provide an approximation of the particle's self-induced flow disturbance. \citet{Esmaily2018} propose a model in which a computational cell is treated as a solid object dragged at a velocity identical to that of the disturbance created by the particle. Their model is extended to wall bounded flows in \cite{Pakseresht2020}. Deriving an analytical closed-form expression for the central component of the regularized Oseenlet operator that is aligned with undisturbed flow, considering a Gaussian filtering kernel, \citet{Balachandar2019} propose a model that accurately predicts the particle's self-induced flow disturbance at finite Reynolds numbers. This work is extended to nonlinear regularizations by \citet{Poustis2019}, and to a compact polynomial filtering kernel \cite{Wendland1995} by \citet{Evrard2020a}, based on the regularized Stokeslet operator introduced by \citet{Cortez2001}. The work of \cite{Balachandar2019} is also generalized to a vector correction procedure in \cite{Balachandar2023}, enabling cases for which the direction of the momentum feedback need not be aligned with the relative velocity of the particle. \citet{Pakseresht2021} propose to solve an auxiliary set of governing equations that extracts all particle-induced velocity perturbations at once. \citet{Horwitz2022} develop a discrete Green's operator for the Stokes equations to provide an estimator of the particle's self-induced flow disturbance that naturally handles boundary conditions. Finally, a correction for finite volume-fraction effects on the particle's self-induced velocity disturbance is proposed by \citet{Kim2024}.\medskip

Models based on steady solutions to the Stokes, Oseen, or Navier-Stokes equations \cite[e.g.,][]{Maxey2001,Ireland2017,Horwitz2018,Evrard2020a,Horwitz2022} can fail to meaningfully predict a particle's self-induced flow disturbance when the time-scale associated with the particle motion is similar to, or smaller than the time-scale at which the flow disturbance develops. Among the models listed in the previous paragraph, few consider the transient evolution of the particle's self-induced flow disturbance \cite{Gualtieri2015,Battista2019,Esmaily2018,Balachandar2019,Pakseresht2020,Pakseresht2021,Balachandar2023}. Even fewer combine this feature with the ability to consider particles moving at finite Reynolds numbers \cite{Esmaily2018,Balachandar2019,Pakseresht2020}, while all relying on some form of empirical fitting. To the authors' knowledge, no model yet presents the ability to concurrently consider transient effects, particles moving at finite Reynolds numbers, and an arbitrary orientation of the momentum feedback with respect to the relative velocity of the particle. This manuscript presents an attempt at doing so, which relies on the linearization of the equations governing the particle's self-induced flow disturbance so as to express its velocity as a temporal and spatial convolution integral involving known analytical operators. The temporal convolution integral is approximated with what is essentially a ``left-hand'' integration rule, while the spatial convolution integral is estimated using pre-computed maps of the (spatially and temporally) regularized transient Stokeslet operators. The resulting model is free of heuristics and empirical coefficients, but requires seeding and keeping track of fluid tracers along the trajectory of a particle, which presents a non-negligeable computational cost. The amount of tracers needing to be tracked for the model to remain accurate, however, can be kept to a minimum owing to the temporal and spatial decay of the transient Stokeslet operators.\medskip

The remainder of this manuscript is organized as follows: Section~\ref{sec:problem_position} introduces the volume-filtering of the incompressible Navier-Stokes equations in the context of particle-laden flow and the terms requiring closure; Section~\ref{sec:model} describes the proposed model for the estimation of a particle's self-induced flow disturbance, used to recover the undisturbed velocity associated with that particle; Section~\ref{sec:special_cases} discusses special cases for which the proposed model recovers well-known analytical solutions; Section~\ref{sec:numerics} details the numerical implementation of the model and studies its spatial and temporal convergence; Section~\ref{sec:test_cases} presents results of VF-EL simulations that employ the proposed model, and verifies its accuracy across a wide range of flow and numerical parameters. It also discusses how to balance computational cost with the accuracy of the model. Finally, conclusions are drawn in Section~\ref{sec:conclusions}.

\section{Problem position}
\label{sec:problem_position}
This section introduces the equations governing particle-laden flows and their volume-filtering. After listing the terms requiring closure, we identify the problem addressed by the model proposed in Section~\ref{sec:model}.

\subsection{Governing equations of incompressible particle-laden flow}
\begin{figure}
  \includegraphics{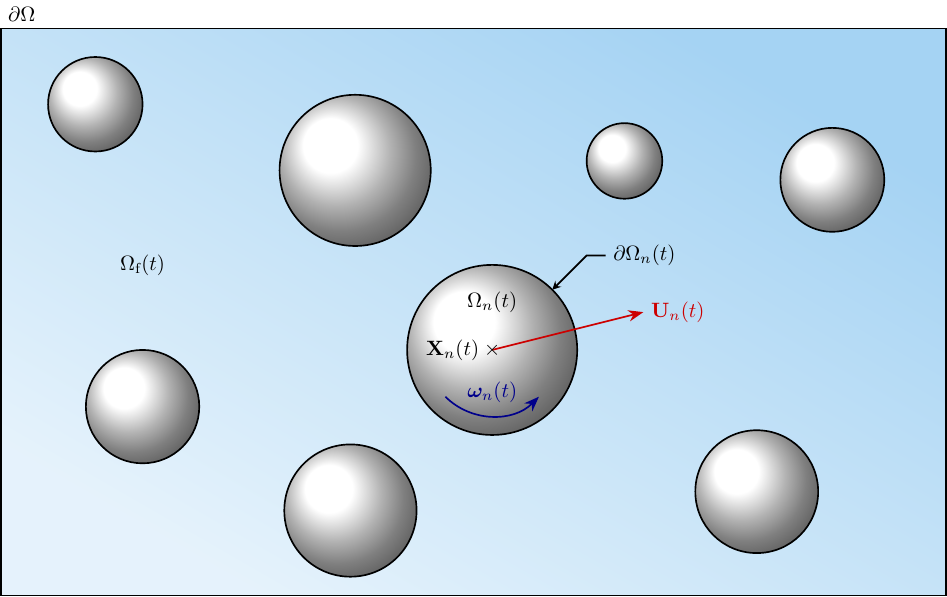}
  \caption{The flow domain $\smash{\Omega}$ is the union of the domain occupied by the fluid, $\smash{\Omega_\text{f}(t)}$, and that occupied by the particles, $\smash{\cup_n \Omega_{n}(t)}$.}\label{fig:problemdescription}
\end{figure}

Let us consider a flow domain $\Omega$ within which $N$ freely moving rigid particles, occupying the volumes $\Omega_{n}(t), n \in \{1,\ldots,N\}$, are surrounded by an incompressible fluid with constant density and viscosity occupying $\smash{\Omega_\text{f}(t) = \Omega \setminus \left(\cup_n \Omega_{n}(t)\right)}$, as illustrated in Figure~\ref{fig:problemdescription}. Gravity is, for now, ignored.
The equations governing the fluid motion inside $\smash{\Omega_\text{f}(t)}$ are the continuity and momentum equations
\begin{align}
  \nabla \cdot \vecu & = 0 \, , \\
  \rho \left(\dfrac{\partial\vecu}{\partial t} + \nabla \cdot \left(\vecu \otimes \vecu\right)\right) & = -\nabla p + \mu\nabla^2\vecu \, ,
\end{align}
where $\smash{\vecu}$ is the velocity vector of the fluid, $\smash{p}$ is the pressure within the fluid, $\smash{\rho}$ is the fluid density, and $\smash{\mu}$ is the dynamic fluid viscosity. At the particle surfaces, the fluid velocity, $\smash{\vecu}$, is subject to the boundary condition
\begin{align}
   \vecu(\mathbf{x},t) & = \vecU_{n}(t) + \boldsymbol{\omega}_{n}(t) \times \left(\vecx-\vecX_{n}(t) \right) \, , \quad \forall  \mathbf{x} \in \partial \Omega_{n}(t) , n \in \{1,\ldots,N\} \, , 
\end{align}
where $\smash{\vecX_{n}}$, $\smash{\vecU_{n}}$ and $\smash{\boldsymbol{\omega}_{n}}$ are the center of mass, translational and rotational velocities of the $n^\mathrm{th}$ particle, respectively.
The equations governing the motion of the $n^\text{th}$ rigid particle are
\begin{align}
  \dfrac{\mathrm{d} \vecX_{n}(t)}{\mathrm{d}t} &= \vecU_{n}(t) \, , &
  M_{n} \dfrac{\mathrm{d} \vecU_{n}(t)}{\mathrm{d}t} &= \vecF_{n}(t) \, , \label{eq:Ppos}\\
  \dfrac{\mathrm{d} \boldsymbol{\theta}_{n}(t)}{\mathrm{d}t} &= \boldsymbol{\omega}_{n}(t) \, ,  & 
  \boldsymbol{{\Lambda}}_{n} \cdot \dfrac{\mathrm{d} \boldsymbol{\omega}_{n}(t)}{\mathrm{d}t} &= \mathbf{T}_{n}(t) \, ,\label{eq:Prot}
\end{align}
where $\smash{M_{n}}$ is the mass of the $\smash{n^\text{th}}$ particle, $\smash{\boldsymbol{{\Lambda}}_{n}}$ its moment of inertia tensor, and $\smash{\boldsymbol{\theta}_{n}}$ its angular position. The vectors $\smash{\vecF_{n}}$ and $\smash{\mathbf{T}_{n}}$ correspond to the resultants of the fluid forces and torques acting on the $\smash{n^\text{th}}$ particle, respectively, given as
\begin{equation}
  \vecF_{n}(t) = \int_{\partial \Omega_{n}(t)} \left(-p(\mathbf{x},t) \mathbf{n} + {\boldsymbol{\tau}}(\mathbf{x},t) \cdot \mathbf{n}\right) \ \mathrm{d}\mathbf{x} \, ,
  \label{eq:Fpn}
\end{equation}
and
\begin{equation}
  \mathbf{T}_{n}(t) = \int_{\partial \Omega_{n}(t)} \left(-p(\mathbf{x},t) \mathbf{n} + {\boldsymbol{\tau}}(\mathbf{x},t) \cdot \mathbf{n}\right) \times \left(\vecx-\vecX_{n}(t)\right) \  \mathrm{d}\mathbf{x} \, ,
\end{equation}
where $\smash{\mathbf{n}}$ is the normal to $\smash{\partial \Omega_{n}(t)}$ pointing towards the fluid, and $\smash{\boldsymbol{\tau}}$ the viscous stress tensor of the fluid.
The system of all previous equations can be closed and solved numerically. However, its accurate numerical solution requires a spatial resolution that is prohibitively expensive as the number of particles becomes significant, since a minimum of $15$ to $20$ computational cells across each particle diameter is necessary~\cite{Cheron2023}. Currently, it is possible to accurately solve such systems with up to $\smash{N \sim \mathcal{O}(10^5)}$ particles \cite[see, e.g.,][]{Uhlmann2017,Schneiders2017,Schneiders2019}, leading to $\smash{\mathcal{O}(10^9)}$ computational degrees of freedom per timestep. For systems with many more than $\smash{N \sim \mathcal{O}(10^5)}$ particles, one must resort to solving an averaged representation of the flow, for which a much less restrictive spatial resolution can be chosen.

\subsection{Volume-filtered Euler-Lagrange modeling}
\begin{figure}\centering
\begin{tabular}{ccc}
  \includegraphics{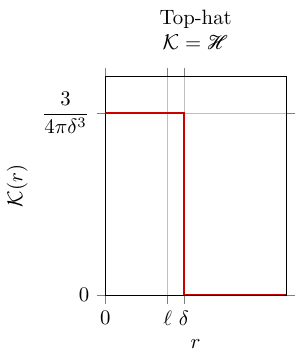} &
  \includegraphics{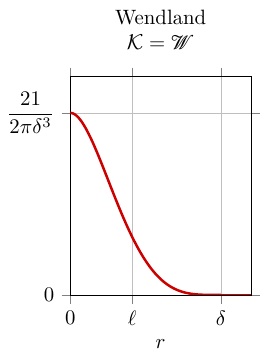} &
  \includegraphics{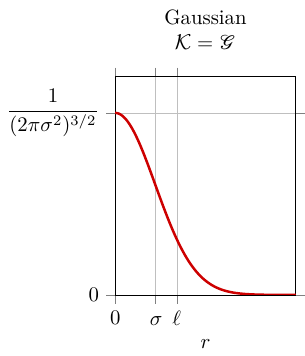}
\end{tabular}  
\caption{Examples of three convolution kernels that can be used for VF-EL simulations: the top-hat kernel $\smash{\mathscr{H}}$, defined in Eq.~\eqref{eq:kerneltophat}, the polynomial Wendland kernel $\smash{\mathscr{W}}$~\cite{Wendland1995}, defined in Eq.~\eqref{eq:defwendland}, and the Gaussian kernel $\smash{\mathscr{G}}$, defined in Eq.~\eqref{eq:kernelgaussian}. The kernels $\smash{\mathscr{H}}$ and $\smash{\mathscr{W}}$ are compactly supported on $\smash{\left[0,\delta\right]}$, whereas $\smash{\mathscr{G}}$ is supported on $\smash{\mathbb{R}^+}$. Each kernel is normalised, \ie satisfies Eq.~\eqref{eq:kernelnormalisation}. The length-scale $\ell$ associated with each kernel, as defined by Eq.~\eqref{eq:defell}, is indicated on the horizontal axes.}
  \label{fig:exampleskernels}
\end{figure}

Volume-averaged or volume-filtered approaches are widely used for studying the behavior of particle-laden flows at the meso- or macro-scale, often referred to as volume-filtered Euler-Lagrange (VF-EL) methods~\cite{Capecelatro2013,Balachandar2019}. They rely on the spatial convolution\footnote{The spatial convolution of two scalar functions $\varphi$ and $\psi$ in $\smash{\mathbb{R}^3}$ is given as $\left[\,\varphi \Conv \psi \,\right] (\mathbf{x},t) = \int_{\mathbb{R}^3} \varphi(\mathbf{y},t) \, \psi(\mathbf{x}-\mathbf{y},t) \ \mathrm{d}{\mathbf{y}}$} of the governing equations with a normalized radial kernel $\K : \mathbb{R}^+ \to \mathbb{R}^+$. In order for this kernel to be normalized, its integral over the three-dimensional real space must satisfy
\begin{equation}
	 \int_{\mathbb{R}^3} \K\left(\left\|\vecx\right\|\right) \, \mathrm{d}{\vecx} = 1 \ . \label{eq:kernelnormalisation}
\end{equation}
The kernel length-scale, $\ell$, is defined as in \cite{Anderson1967} as the radius of the ball\footnote{The ball of radius $\ell$, in $\smash{\mathbb{R}^3}$, is the volume bounded by the sphere of radius $\ell$.} $\smash{\mathbb{B}_\ell}$ for which
\begin{equation}
	\int_{\mathbb{B}_\ell} \K\left(\left\|\vecx\right\|\right) \, \mathrm{d}{\vecx} = 4\pi \int_0^\ell r^2 \K\left(r\right) \, \mathrm{d}r = \frac{1}{2} \ . \label{eq:defell}
\end{equation}
Examples of kernels that can be employed for VF-EL simulations are provided in Figure~\ref{fig:exampleskernels}. 
The resulting volume-filtered governing equations, which are now defined on the entire domain $\smash{\Omega}$ instead of being limited to the fluid domain $\smash{\Omega_\mathrm{f}(t)}$, read as \citep{Anderson1967,Capecelatro2013,Balachandar2019}
\begin{align}
	\frac{\partial \vfrac}{\partial t} + \nabla \cdot \bar{\vecu}_\vfrac & = 0 \, , \label{eq:vfcont}\\
	\rho \left( \frac{\partial \bar{\vecu}_\vfrac}{\partial t} + \nabla \cdot \left(\bar{\vecu}_\vfrac \otimes \bar{\vecu}_\vfrac \right)\right) & = -\nabla \bar{p}_\vfrac + \mu \nabla^2 \bar{\vecu}_\vfrac + \vecf - \boldsymbol{\mathscr{U}}_{\!\mathrm{Re}} + \boldsymbol{\mathscr{U}}_{\!\mu} \, . \label{eq:vfmom}
\end{align}
In these volume-filtered equations:
\begin{itemize}
\item $\smash{\vfrac}$ is the fluid volume fraction, defined as
\begin{equation}
	\vfrac = \K \Conv \zeta_\mathrm{f} \, , \label{eq:defvfrac}
\end{equation}
with $\smash{\zeta_\mathrm{f}}$ the fluid indicator function \citep{Anderson1967,Drew1983,Ishii2006},
\begin{equation}
\zeta_\mathrm{f} (\vecx,t)= \left\{\begin{array}{ll} 1 & \vecx \in \Omega_\text{f}(t) \\ 0 & \vecx \notin \Omega_\text{f}(t) \end{array}\right. \, . 
\end{equation}
The fluid volume fraction can equivalently be written as
\begin{equation}
	\vfrac = 1 - \sum_{n=1}^{N} \vfrac_n  \, , \label{eq:defvfrac2}
\end{equation}
where $\smash{\vfrac_n}$ is the volume fraction contribution of the $n^\mathrm{th}$ particle given as
\begin{equation}
  \vfrac_n = \K \Conv \zeta_n \, , \label{eq:defvfracn}
\end{equation}
with $\smash{\zeta_n}$ the indicator function associated with the $n^\mathrm{th}$ particle,
\begin{equation}
\zeta_n (\vecx,t)= \left\{\begin{array}{ll} 1 & \vecx \in \Omega_n(t) \\ 0 & \vecx \notin \Omega_n(t) \end{array}\right. \, .
\end{equation}
The fluid volume fraction, $\smash{\vfrac}$, is defined over the entire domain $\Omega$. By construction with Eq.~\eqref{eq:defvfrac} or Eq.~\eqref{eq:defvfrac2}, it satisfies $\smash{0\le\vfrac\le1}$. If the radius of the kernel's support is larger than the radius of the particles, one can even write the strict inequality $\smash{0<\vfrac\le1}$; a necessary condition for the volume-filtered governing equations not to become ill-posed. The convolution integral in Eq.~\eqref{eq:defvfracn} can be derived into a closed form expression for some specific kernels. \citet{Balachandar2023} have for instance shown that when $\smash{\K}$ is the Gaussian kernel $\smash{\mathscr{G}}$ of standard deviation $\sigma$ (defined in Eq.~\eqref{eq:kernelgaussian}), the volume fraction contribution of the $n^\mathrm{th}$ particle is exactly given as
\begin{align}
	\vfrac_n (\vecx,t) & = \left[\,\mathscr{G} \Conv \zeta_n\,\right] (\vecx,t) \\
  & = \dfrac{2\pi\sigma^4}{\chi_n(\vecx,t)}
  \left(\mathscr{G}\left(\chi_n(\vecx,t)+r_n\phantom{\sqrt{2}\!\!\!\!\!\!\!\!}\right) - \mathscr{G}\left(\chi_n(\vecx,t)-r_n\phantom{\sqrt{2}\!\!\!\!\!\!\!\!}\right)\right) \nonumber \\ 
  & \quad\quad + \dfrac{1}{2}\left(\erf\left(\left(\chi_n(\vecx,t)+r_n\right)/\sqrt{2}\sigma\right) - \erf\left(\left(\chi_n(\vecx,t)-r_n\right)/\sqrt{2}\sigma\right)\right) \label{eq:defvfracgaussian}
\end{align}
with $\smash{r_n}$ the radius of the $n^\mathrm{th}$ particle, and $\smash{\chi_n(\vecx,t) = \|\vecx-\vecX_n(t)\|}$ the distance to its center. 
In practice, $\smash{\vfrac_n}$ is often approximated as
\begin{equation}
	\vfrac_n(\vecx,t) \simeq  \K\left(\chi_n(\vecx,t)\right) V_n \, ,
\end{equation}
with $\smash{V_n}$ the volume of the $\smash{n^\mathrm{th}}$ particle, under the assumption that the length-scale of the filter, $\ell$, is large compared to the particle radius~$r_n$. With this approximation, $\smash{\vfrac}$ is not bounded by construction anymore, and one must make sure that $\ell$ is large enough as to guarantee that $\smash{0<\vfrac\le1}$~\cite{Evrard2019}. 

\item $\smash{\bar{\vecu}_\vfrac}$ and $\smash{\bar{p}_\vfrac}$ are the $\vfrac$-weighted filtered velocity and pressure, defined as
\begin{align}
	\bar{\vecu}_\vfrac & = \K \Conv \zeta_\vecu = \vfrac \bar{\vecu}\, , \\ \bar{p}_\vfrac & = \K \Conv \zeta_p= \vfrac \bar{p}\ ,
\end{align}
with
\begin{align}
\zeta_\vecu  (\vecx,t) &= \left\{\begin{array}{ll} \vecu(\vecx,t) & \vecx \in \Omega_\text{f}(t) \\ 0 & \vecx \notin \Omega_\text{f}(t) \end{array}\right.
\, , \\
\zeta_p (\vecx,t) &=\left\{\begin{array}{ll} p(\vecx,t) & \vecx \in \Omega_\text{f}(t) \\ 0 & \vecx \notin \Omega_\text{f}(t) \end{array}\right. \ .
\end{align}
Note that these quantities, as well as the filtered velocity and pressure, $\smash{\bar{\vecu}}$ and $\smash{\bar{p}}$, are defined over the entire domain $\Omega$.
\item $\smash{\vecf}$ represents the transfer of momentum between particles and fluid given as
\begin{equation}
	\vecf(\vecx,t) = -\sum_{n=1}^{N} \int_{\partial \Omega_{n}(t)} \K\left(\left\|\vecx-\vecy\right\|\right) \left(-p(\vecy,t) \mathbf{n} + {\boldsymbol{\tau}}(\vecy,t) \cdot \mathbf{n}\right) \ \mathrm{d}\vecy  \, .
\end{equation}
It is often approximated as
\begin{equation}
	\vecf(\vecx,t) \simeq -\sum_{n=1}^{N} \K\left(\left\|\vecx-\vecX_n(t)\right\|\right) \vecF_n(t) \, ,
\end{equation}
where $\smash{\vecF_n}$ is the force exerted by the fluid on the $n^\mathrm{th}$ particle, previously defined in Eq.~\eqref{eq:Fpn}, although this approximation is only strictly valid for $\ell \gg r_n$~\cite{Hausmann2024}.
\item $\smash{\boldsymbol{\mathscr{U}}_{\!\mathrm{Re}}}$ is an unclosed term akin to the divergence of Reynolds stresses, given by
\begin{equation}
	\boldsymbol{\mathscr{U}}_{\!\mathrm{Re}} = \nabla \cdot \left(\vfrac \overline{\vecu \otimes \vecu} - \bar{\vecu}_\vfrac \otimes \bar{\vecu}_\vfrac\right) \ .
\end{equation}
\item $\smash{\boldsymbol{\mathscr{U}}_{\!\mu}}$ an unclosed term resulting from the filtering of the viscous stresses, given by
\begin{equation}
	\boldsymbol{\mathscr{U}}_{\!\mu} = \mu \left( \vfrac \overline{\nabla^2 \vecu} - \nabla^2 \bar{\vecu}_\vfrac \right) \ .
\end{equation}
\end{itemize}

\begin{figure}[h]
  \includegraphics{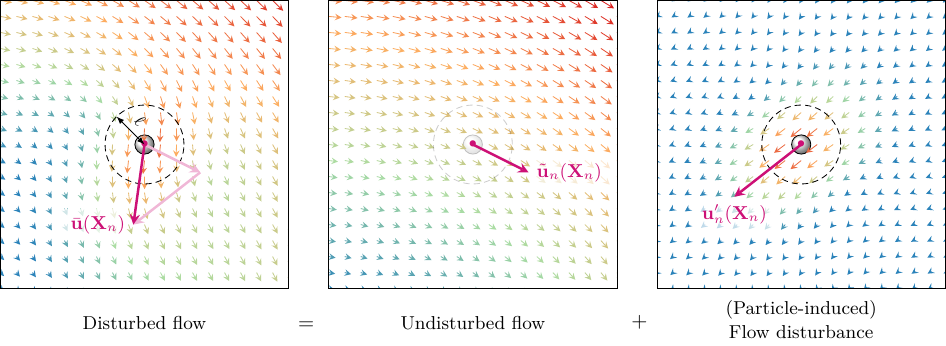}
  \caption{The velocity field solution to the volume-filtered governing equations \eqref{eq:vfcont} and \eqref{eq:vfmom} (the \textit{disturbed flow} velocity) is equal to the sum of the \textit{undisturbed flow} velocity for a given particle and of the \textit{velocity disturbance} induced by that particle. The undisturbed flow velocity is needed to accurately approximate the fluid force acting on the particle using the MRG equation.}
  \label{fig:undistflow}
\end{figure}

In this new set of volume-filtered governing equations, and from now on, we neglect particle rotation. The motion of the particles is still governed by Eqs.~\eqref{eq:Ppos}, however $\smash{\vecF_{n}(t)}$ cannot be estimated using Eq.~\eqref{eq:Fpn} since the quantities that are solved for and known are the volume-filtered pressure and velocity, $\smash{\bar{p}}$ and $\smash{\bar{\vecu}}$, in lieu of the actual pressure and velocity of the flow around the particle, $\smash{{p}}$ and $\smash{{\vecu}}$. Reduced models must therefore be used to estimate the fluid force acting on a particle, as given for instance by the Maxey-Riley-Gatignol (MRG) equation \citep{Maxey1983,Gatignol1983},
\begin{equation}
\begin{aligned}
  \mathbf{F}_{n}(t) & = \mathbf{F}_{n,\text{drag}}\left(\tilde{\vecu}_{n,\partial \Omega_{n}}(t),\mathbf{U}_{n}(t)\right) \\
  & \quad\quad + \mathbf{F}_{n,\text{undist.}}\left(\tilde{\vecu}_{n,\Omega_{n}}(t)\right) \\
  & \quad\quad\quad + \mathbf{F}_{n,\text{virt.}}\left(\tilde{\vecu}_{n,\Omega_{n}}(t),\mathbf{U}_{n}(t)\right) \\
  & \quad\quad\quad\quad + \mathbf{F}_{n,\text{hist.}}\left(\tilde{\vecu}_{n,\partial \Omega_{n}}(t),\mathbf{U}_{n}(t)\right) \, .
\end{aligned}
\label{eq:mrg}
\end{equation}
In this equation, $\smash{\mathbf{F}_{n,\text{drag}}}$ is the steady viscous drag force contribution, $\smash{\mathbf{F}_{n,\text{undist.}}}$ is the force contribution of the undisturbed flow stresses, $\smash{\mathbf{F}_{n,\text{virt.}}}$ is the unsteady virtual mass force contribution, and $\smash{\mathbf{F}_{n,\text{hist.}}}$ is the unsteady viscous history force contribution. 
The variable $\smash{\tilde{\vecu}_n}$ corresponds to the \textit{undisturbed fluid flow velocity for the $n^\text{th}$ particle}, which is the fluid velocity as though the $n^\text{th}$ particle under consideration had been taken out of the flow domain (see Figure~\ref{fig:undistflow}).\medskip

The subscripts $\smash{\partial \Omega_{n}}$ and $\smash{\Omega_{n}}$ in the MRG Eq.~\eqref{eq:mrg} indicate that the undisturbed velocity has been averaged over the particle's surface and volume, respectively, i.e.,
\begin{align}
  \tilde{\vecu}_{n,\partial \Omega_{n}}(t) & = \frac{1}{S_{n}} \int_{\partial \Omega_{n}(t)} \tilde{\vecu}_n(\mathbf{x},t) \ \mathrm{d}\mathbf{x} \, , \\
  \tilde{\vecu}_{n,\Omega_{n}}(t) & = \frac{1}{V_{n}} \int_{\Omega_{n}(t)} \tilde{\vecu}_n(\mathbf{x},t) \ \mathrm{d}{\vecx} \, ,
\end{align}
where $\smash{S_{n}}$ and $\smash{V_{n}}$ are the surface area and volume of the $n^\text{th}$ particle. Assuming spherical particles and from the Taylor series expansion of the undisturbed velocity field inside $\smash{\Omega_{n}}$, these averaged quantities can be approximated as \citep{Faxen1922}
\begin{align}
  \tilde{\vecu}_{n,\partial \Omega_{n}}(t) & \simeq \tilde{\vecu}_n(\vecX_{n}(t),t) + \frac{r_n^2}{6}\nabla^2 \tilde{\vecu}_n(\vecX_{n}(t),t) \, , \\
  \tilde{\vecu}_{n,\Omega_{n}}(t) & \simeq \tilde{\vecu}_n(\vecX_{n}(t),t) + \frac{r_n^2}{10}\nabla^2 \tilde{\vecu}_n(\vecX_{n}(t),t) \, .
\end{align}

\subsection{Objectives of this manuscript}
In order to make rigorous and optimal use of the volume-filtered Euler-Lagrange method, one must:
\begin{enumerate}
	\item Close the terms $\smash{\boldsymbol{\mathscr{U}}_{\!\mathrm{Re}}}$ and $\smash{\boldsymbol{\mathscr{U}}_{\!\mu}}$;
	\item Close the momentum transfer term $\smash{\vecf}$, which requires to estimate the undisturbed fluid velocity $\smash{\tilde{\vecu}}$ and its Laplacian $\smash{\nabla^2\tilde{\vecu}}$ at the location of each particle from the knowledge of the filtered flow variables.
\end{enumerate}
In this manuscript, we shall focus on addressing the second of these two tasks, particularly focusing on recovering an accurate undisturbed velocity from the filtered flow. To that end, we propose a new model for estimating the velocity disturbance induced by the $n^\mathrm{th}$ particle, $\smash{\vecu^\prime_n}$, such that its undisturbed velocity can be recovered as $\smash{\tilde{\vecu}_n = \bar{\vecu} - \vecu^\prime_n}$. For insights on how to close the term $\smash{\boldsymbol{\mathscr{U}}_{\!\mathrm{Re}}}$, an analytical closure for the term $\smash{\boldsymbol{\mathscr{U}}_{\!\mu}}$, as well as a discussion on the regularization of the momentum exchange term $\vecf$, we refer the reader to the recent work of \citet{Hausmann2024}.

\section{Model description}
\label{sec:model}
In this section, a model for recovering the undisturbed flow velocity and its Laplacian at the location of each particle is derived in the context of volume-filtered Euler-Lagrange modeling. This model relies on the linearization of the equations governing a particle's self-induced flow disturbance, which are solved using Green's functions for the resulting linear operators and discretizing the corresponding convolution integrals. 

\subsection{Linearized governing equations of the particle's self-induced flow disturbance}
\label{sec:Linearizedisteq}
Let us consider the $n^\mathrm{th}$ particle of our ensemble of $N$ particles, and rewrite the volume-filtered governing equations of the flow in the frame of reference of this particle. The position vector, in this frame of reference, is expressed as
\begin{equation}
	\vecy = \vecx - \vecX_n(t) \, ,
\end{equation}
\ie, relative to the center of the $n^\mathrm{th}$ particle. Moreover, we neglect the unclosed terms $\smash{\boldsymbol{\mathscr{U}}_{\!\mathrm{Re}}}$ and $\smash{\boldsymbol{\mathscr{U}}_{\!\mu}}$. The resulting volume-filtered governing equations read as
\begin{align}
	\frac{\partial \vfrac}{\partial t} + \nabla \cdot \bar{\vecv}_\vfrac & = 0 \, , \label{eq:vfumcont}\\
	\rho \left( \frac{\partial \bar{\vecv}_\vfrac}{\partial t} + \nabla \cdot \left(\bar{\vecv}_\vfrac \otimes \bar{\vecv}_\vfrac \right)\right) & = -\dfrac{\mathrm{d}\vecU_n}{\mathrm{d}t} -\nabla \bar{p}_\vfrac + \mu \nabla^2 \bar{\vecv}_\vfrac  - \sum_{m=1}^{N} \K\left(\left\|\vecy-\left(\vecX_m-\vecX_n\right)\right\|\right) \vecF_m  \, ,\label{eq:vfummom}
\end{align}
where $\smash{\bar{\vecv}_\vfrac}$ is the $\vfrac$-weighted filtered fluid velocity relative to that of the $n^\mathrm{th}$ particle,
\begin{equation}
  \bar{\vecv}_\vfrac(\vecy,t) = \bar{\vecu}_\vfrac(\vecx,t) - \vecU_n(t) \, .
\end{equation}
The undisturbed flow for the $n^\mathrm{th}$ particle is the conceptual flow in which the $n^\mathrm{th}$ particle has been removed from the flow domain. The volume-filtered equations governing this undisturbed flow thus are
\begin{align}
	\frac{\partial \left(\vfrac+\vfrac_n\right)}{\partial t} + \nabla \cdot \tilde{\vecv}_{n,\vfrac} & = 0 \, , \label{eq:vfutildemcont}\\
	\rho \left( \frac{\partial \tilde{\vecv}_{n,\vfrac}}{\partial t} + \nabla \cdot \left(\tilde{\vecv}_{n,\vfrac} \otimes \tilde{\vecv}_{n,\vfrac} \right)\right) & =  -\dfrac{\mathrm{d}\vecU_n}{\mathrm{d}t} -\nabla \tilde{p}_{n,\vfrac} + \mu \nabla^2 \tilde{\vecv}_{n,\vfrac} +\K\left(\left\|\vecy\right\|\right) \vecF_n \label{eq:vfutildemmom}\\ 
	& \quad\quad\quad\quad\quad\quad\quad\quad\quad\quad -\sum_{m=1}^{N} \K\left(\left\|\vecy-\left(\vecX_m-\vecX_n\right)\right\|\right) \vecF_m \, , \nonumber
\end{align}
where $\smash{\vfrac_n = \mathscr{G} \Conv \zeta_n}$ is the volume fraction contribution of the $n^\mathrm{th}$ particle, introduced in Eq.~\eqref{eq:defvfracn} and satisfying $\smash{0\le\vfrac_n \le 1-\vfrac}$. The quantity $\smash{\tilde{\vecv}_{n,\vfrac} = \left(\vfrac+\vfrac_n\right)\tilde{\vecv}_n}$ is the $\vfrac$-weighted filtered undisturbed velocity, whereas $\smash{\tilde{\vecv}_n}$ is the filtered undisturbed velocity for the $n^\mathrm{th}$ particle. Note that in the case of an isolated particle situated far away from other particles, $\smash{\vfrac + \vfrac_n \simeq 1}$ and therefore $\smash{\smash{\tilde{\vecv}_n} \simeq \tilde{\vecv}_{n,\vfrac}}$ locally. We now define the $\vfrac$-weighted filtered velocity disturbance induced by the $n^\mathrm{th}$ particle as
\begin{equation}
	{\vecv}^\prime_{n,\vfrac} 
  = \bar{\vecv}_{\vfrac} - \tilde{\vecv}_{n,\vfrac} 
  \, . \label{eq:vfracdisturbance}
\end{equation}
Note that since $\smash{\bar{\vecv}_{\vfrac}}$ and $\smash{\tilde{\vecv}_{n,\vfrac}}$ are both expressed relatively to the velocity of the $n^\mathrm{th}$ particle, then
\begin{align}
	{\vecv}^\prime_{n,\vfrac}(\vecy,t) & = \bar{\vecv}_{\vfrac}(\vecy,t) - \tilde{\vecv}_{n,\vfrac}(\vecy,t) \nonumber \\
  & = \bar{\vecv}_{\vfrac}(\vecy,t) + \vecU_n(t) - \left(\tilde{\vecv}_{n,\vfrac}(\vecy,t)  + \vecU_n(t)\right) \nonumber \\
  & = \bar{\vecu}_{\vfrac}(\vecx,t) - \tilde{\vecu}_{n,\vfrac}(\vecx,t)  \nonumber \\
  & = {\vecu}^\prime_{n,\vfrac}(\vecx,t) \, .
\end{align}

Subtracting the volume-filtered governing equations of the undisturbed flow, Eqs.~\eqref{eq:vfutildemcont} and \eqref{eq:vfutildemmom}, from those of the disturbed flow, Eqs.~\eqref{eq:vfumcont} and \eqref{eq:vfummom}, we obtain governing equations for the $\vfrac$-weighted velocity and pressure disturbances,
\begin{align}
	-\frac{\partial \vfrac_n}{\partial t} + \nabla \cdot {\vecv}^\prime_{n,\vfrac} & = 0 \, , \\
	\rho \left( \frac{\partial {\vecv}^\prime_{n,\vfrac}}{\partial t} + \nabla \cdot \left(\tilde{\vecv}_{n,\vfrac} \otimes {\vecv}^\prime_{n,\vfrac} + {\vecv}^\prime_{n,\vfrac} \otimes \bar{\vecv}_{\vfrac} \right)\right) & = -\nabla {p}^{\prime}_{n,\vfrac} + \mu \nabla^2 {\vecv}^\prime_{n,\vfrac} - \K\left(\left\|\vecy\right\|\right) \vecF_n \, .
\end{align}
Since we are in the frame of reference of the $n^\mathrm{th}$ particle, $\smash{\partial \vfrac_n/\partial t \equiv 0}$, so the equations simplify to
\begin{align}
	\nabla \cdot {\vecv}^\prime_{n,\vfrac} & = 0 \, , \\
	\rho \left( \frac{\partial {\vecv}^\prime_{n,\vfrac}}{\partial t} + \tilde{\vecv}_{n,\vfrac} \cdot \nabla {\vecv}^\prime_{n,\vfrac} + {\vecv}^\prime_{n,\vfrac} \left(\nabla \cdot \tilde{\vecv}_{n,\vfrac}\right) +  {\vecv}^\prime_{n,\vfrac} \cdot \nabla \bar{\vecv}_{\vfrac}\right) & = -\nabla {p}^{\prime}_{n,\vfrac} + \mu \nabla^2 {\vecv}^\prime_{n,\vfrac} - \K\left(\left\|\vecy\right\|\right) \vecF_n \label{eq:distmomnonsimplified} 
  \, . 
\end{align}
We further simplify these equations by neglecting the last two advection terms on the left-hand side of Eq.~\eqref{eq:distmomnonsimplified}, assuming that the Reynolds number associated with the velocity disturbance is vanishingly small, \ie, $\smash{\rho \ell \|{\vecv}^\prime\| / \mu \ll 1}$. The resulting approximate governing equations of the particle-induced flow disturbance, in the frame of reference of the $n^\mathrm{th}$ particle, are linear and read as
\begin{align}
	\nabla \cdot {\vecv}^\prime_{n,\vfrac} & = 0 \, , \\
	\rho \left( \frac{\partial {\vecv}^\prime_{n,\vfrac}}{\partial t} + \tilde{\vecv}_{n,\vfrac} \cdot \nabla {\vecv}^\prime_{n,\vfrac}\right) & = -\nabla {p}^{\prime}_{n,\vfrac} + \mu \nabla^2 {\vecv}^\prime_{n,\vfrac} - \K\left(\left\|\vecy\right\|\right) \vecF_n \, .
\end{align}
Finally, in order to derive an approximate analytical solution to this system of equations, we assume that $\smash{\tilde{\vecv}_{n,\vfrac}}$ varies little over the length-scales associated with the particle-induced flow disturbance. We can then consider that $\smash{\tilde{\vecv}_{n,\vfrac} (\vecy,t) \simeq \tilde{\mathbf{V}}_{n,\vfrac}}(t)$ in the relative vicinity of $\smash{\vecy = \boldsymbol{0}}$. In any new coordinate system centered around $\smash{\tilde{\vecY}_n(t)}$, solution to 
\begin{equation}
	\frac{\mathrm{d}\tilde{\vecY}_n(t)}{\mathrm{d}t} = \tilde{\mathbf{V}}_{n,\vfrac}(t) \, ,
\end{equation}
and subject to some initial conditions (that we do not yet specify), the position vector reads as
\begin{equation}
	\vecz = \vecy - \tilde{\vecY}_n(t) \, , 
\end{equation}
and we can rewrite the volume-filtered governing equations of the undisturbed flow as
\begin{align}
	\nabla \cdot {\vecw}^\prime_{n,\vfrac} & = 0 \, , \label{eq:distcont0} \\
	\rho \frac{\partial {\vecw}^\prime_{n,\vfrac}}{\partial t} & = -\dfrac{\mathrm{d}\tilde{\mathbf{V}}_{n,\vfrac}}{\mathrm{d}t} -\nabla {p}^{\prime}_{n,\vfrac} + \mu \nabla^2 {\vecw}^\prime_{n,\vfrac} - \K(\|\vecz + \tilde{\vecY}_n\|) \vecF_n \, , \label{eq:distmom0}
\end{align}
with $\smash{{\vecw}^\prime_{n,\vfrac} (\vecz,t) = {\vecv}^\prime_{n,\vfrac}(\vecy,t) = {\vecu}^\prime_{n,\vfrac}(\vecx,t)}$.
We further choose to neglect the acceleration of this coordinate system, effectively approximating $\smash{{\vecw}^\prime_{n,\vfrac}}$ as the solution to the unsteady Stokes equations
\begin{align}
	\nabla \cdot {\vecw}^\prime_{n,\vfrac} & = 0 \, , \label{eq:distcont} \\
	\rho \frac{\partial {\vecw}^\prime_{n,\vfrac}}{\partial t} & = -\nabla {p}^{\prime}_{n,\vfrac} + \mu \nabla^2 {\vecw}^\prime_{n,\vfrac} - \K(\|\vecz + \tilde{\vecY}_n\|) \vecF_n \, . \label{eq:distmom}
\end{align}
Althought it might seem that we have made a lot of stringent assumptions to reach the system of equations~\eqref{eq:distcont}--\eqref{eq:distmom}, we shall see that its solution does a remarkably good job at approximating the transient evolution of the self-induced particle disturbance, even at finite particle Reynolds numbers.

\subsection{Green's function for the unsteady Stokes equations}
\label{sec:greenvel}
Green's function for the set of linear equations~\eqref{eq:distcont} and \eqref{eq:distmom} is classically referred to as the \textit{transient Stokeslet} operator and given as~\cite{Pozrikidis2011}
\begin{equation}
	\vecg(\vecx,t) = \frac{1}{\mu} \left(\identity \nabla^2 - \nabla \nabla \right) {\B}(\|\vecx\|,t)\, ,
	\label{eq:unsteadyStokeslet1}
\end{equation}
where 
\begin{equation}
	{\B}(r,t) = \frac{1}{8\pi} \left[\sqrt{\frac{4\nu}{\pi t}} - \frac{2\nu}{r} \erf \left(\frac{r}{\sqrt{4\nu t}}\right)\right] \, ,
\end{equation}
and with $\smash{\nu=\mu/\rho}$ the kinematic viscosity of the fluid. Owing to the radial symmetry of $\B$, Eq.~\eqref{eq:unsteadyStokeslet1} also reads as
\begin{equation}
	\vecg(\vecx,t) = \frac{1}{\mu} \left(\identity{\hh}_1(\|\vecx\|,t) + \left(\vecx \otimes \vecx\right) {\hh}_2(\|\vecx\|,t) \right) \, , \label{eq:green2}
\end{equation}
where
\begin{align}
	{\hh}_1(r,t) & = 
	\frac{1}{8\pi r}\left[  \frac{r^2+2\nu t}{rt\sqrt{\pi \nu t}} \exp\left(-\frac{r^2}{4\nu t}\right) - \frac{2\nu}{r^2} \erf \left(\frac{r}{\sqrt{4\nu t}}\right) \right] \, , \\
	{\hh}_2(r,t) & = 
	\frac{1}{8\pi r^3}\left[ - \frac{r^2+6\nu t}{rt\sqrt{\pi \nu t}} \exp\left(-\frac{r^2}{4\nu t}\right) + \frac{6\nu}{r^2} \erf \left(\frac{r}{\sqrt{4\nu t}}\right) \right] \, .
\end{align}
The functions $\smash{{\hh}_1}$ and $\smash{{\hh}_2}$ are singular both in space, at $\smash{r = \|\vecx\| = {0}}$, and in time, at $t=0$. The tensorial operator~$\vecg$ governs the flow response to a point-force impulse in the Stokes regime.

\subsection{Approximate solution of the particle's self-induced velocity disturbance}
\label{sec:modeldist}
The solution to Eqs.~\eqref{eq:distcont} and \eqref{eq:distmom} can be obtained analytically via convolution in space and time of the Green's function given in Eq.~\eqref{eq:green2} with the regularized momentum source in Eq.~\eqref{eq:distmom}, leading to
\begin{equation}
	{\vecw}^\prime_{n,\vfrac} (\vecz,t) = \int_{\mathbb{R}^3} \int_0^t -\K(\|\vecr + \tilde{\vecY}_n(\tau)\|) \vecF_n(\tau) \cdot \vecg(\vecz-\vecr,t-\tau)    \,\mathrm{d}\tau \, \mathrm{d}{\vecr} \, .
\end{equation}
Note that, in this equation, $\smash{\vecr}$ and $\tau$ are spatial and temporal integration variables, respectively.
From the definition of the previously introduced coordinate change, we know that
\begin{equation}
  {\vecw}^\prime_{n,\vfrac} (\vecz,t) = {\vecw}^\prime_{n,\vfrac} (\vecy-\tilde{\vecY}_n(t),t) = {\vecv}^\prime_{n,\vfrac} (\vecy,t) \, ,
\end{equation}
therefore
\begin{equation}
  {\vecv}^\prime_{n,\vfrac} (\vecy,t) = \int_{\mathbb{R}^3} \int_0^t -\K(\|\vecr + \tilde{\vecY}_n(\tau)\|) \vecF_n(\tau) \cdot \vecg(\vecy-\tilde{\vecY}_n(t)-\vecr,t-\tau) \,\mathrm{d}\tau \, \mathrm{d}{\vecr}\, .
\end{equation}
With the variable change $\smash{\vecq = \vecr + \tilde{\vecY}_n(t)}$, this result also reads as
\begin{equation}
  {\vecv}^\prime_{n,\vfrac} (\vecy,t) = \int_{\mathbb{R}^3} \int_0^t -\K(\|\vecq - (\tilde{\vecY}_n(t)-\tilde{\vecY}_n(\tau))\|) \vecF_n(\tau) \cdot \vecg(\vecy-\vecq,t-\tau) \,\mathrm{d}\tau \, \mathrm{d}{\vecq}\, .
\end{equation}
Consistently with the time discretization of the volume-filtered governing equations, we can split $\smash{[0,t]}$ into $K$ time intervals $\smash{[\tk,\tkk]}$ and produce the following first-order approximation of the previous integral
\begin{equation}
	{\vecv}^\prime_{n,\vfrac} (\vecy,t) = \sum_{k=1}^{K} \int_{\mathbb{R}^3} \left[ \K\left(\left\|\vecq - (t-\tk)\tilde{\vecv}_{n,\vfrac}\right\|\right)\vecF_n\left(\tk\right) \cdot \int_{\tkk}^{\tk} \vecg(\vecy-\vecq,t-\tau) \ \mathrm{d}\tau \right] \, \mathrm{d}{\vecq}  \, . \label{eq:model2}
\end{equation}
The time integral in Eq.~\eqref{eq:model2} can now be derived analytically. Integrating the transient Stokeslet operator from $0$ to $t$, we indeed obtain the persistent transient Stokeslet operator
\begin{equation}
	\integral{\vecG}(\vecx,t) = \int_{0}^{t} {\vecg}(\vecx,s) \, \mathrm{d}s = \frac{1}{\mu} \left(\identity\integral{\Hh}_1(\|\vecx\|,t) + \left(\vecx \otimes \vecx\right) \, \integral{\Hh}_2(\|\vecx\|,t) \right) \, ,
\end{equation}
with 
\begin{align}
	\integral{\Hh}_1(r,t) & =\int_{0}^{t} {\hh}_1(r,s) \, \mathrm{d}s= \frac{1}{8\pi r}\left[ 1 + \frac{2}{r}\sqrt{\frac{\nu t}{\pi}}\exp\left( -\frac{r^2}{4\nu t} \right) - \left(1+\frac{2\nu t}{r^2}\right) \erf \left( \frac{r}{\sqrt{4\nu t}}\right) \right] \, ,\\
	\integral{\Hh}_2(r,t) & = \int_{0}^{t} {\hh}_2(r,s) \, \mathrm{d}s= \frac{1}{8\pi r^3}\left[ 1 - \frac{6}{r} \sqrt{\frac{\nu t}{\pi}}\exp\left( -\frac{r^2}{4\nu t} \right) - \left(1 -\frac{6\nu t}{r^2}\right) \erf \left( \frac{r}{\sqrt{4\nu t}}\right)\right] \, .
\end{align}
Note that $\smash{\integral{\Hh}_1}$ and $\smash{\integral{\Hh}_2}$ are only singular in space at $\smash{r = \|\vecx\| = {0}}$.\medskip

The $\vfrac$-weighted velocity disturbance, $\smash{{\vecv}^\prime_{n,\vfrac}}$, can thus be approximated as
\begin{equation}
	{\vecv}^\prime_{n,\vfrac} (\vecy,t) = \sum_{k=1}^{K} \vecF_n\left(\tk\right) \cdot \left[  \int_{\mathbb{R}^3}  \K\left(\left\|\vecq - (t-\tk)\tilde{\vecv}_{n,\vfrac}\right\|\right) \integral{\vecG}(\vecy-\vecq,t-\tau)  \, \mathrm{d}{\vecq} \right]_{\tau\,=\,\tkk}^{\tau\,=\,\tk} \, . \label{eq:model3}
\end{equation}
The integral in Eq.~\eqref{eq:model3} corresponds to the spatial convolution of $\smash{\integral{\vecG}}$, Green's function for the Stokes flow induced by a time-persistent point-source,  with the filtering kernel $\smash{\K}$ centered at the location $\smash{(t-\tk)\tilde{\vecv}_{n,\vfrac}}$. Let us then introduce $\smash{\integral{\vecG}_\K}$, Green's function for the Stokes flow induced by a time-persistent point-source regularized with the kernel $\smash{\K}$,
\begin{equation}
	\integral{\vecG}_\K = \K \Conv \integral{\vecG} \, .
\end{equation}
We also introduce $\smash{\tilde{\vecY}_{n,k}}$, the solution to the transport equation
\begin{equation}
	\frac{\mathrm{d}\tilde{\vecY}_{n,k}(s)}{\mathrm{d}s} = \tilde{\vecv}_{n,\vfrac}(\tilde{\vecY}_{n,k}(s),\tk+s) \, ,
\end{equation}
subject to the initial condition $\tilde{\vecY}_{n,k}(0) = \boldsymbol{0}$. The $\vfrac$-weighted velocity disturbance then reads as
\begin{equation}
	{\vecv}^\prime_{n,\vfrac} (\vecy,t) = \sum_{k=1}^{K} \vecF_n\left(\tk\right) \cdot \left[  \integral{\vecG}_\K\left(\vecy-\tilde{\vecY}_{n,k}(t-\tk),t-\tau\right) \right]_{\tau\,=\,\tkk}^{\tau\,=\,\tk} \, . \label{eq:model4}
\end{equation}
Moving back to the original canonical frame of reference, it follows that
\begin{equation}
	{\vecu}^\prime_{n,\vfrac} (\vecx,t) = \sum_{k=1}^{K} \vecF_n\left(\tk\right) \cdot \left[  \integral{\vecG}_\K\left(\vecx-\tilde{\vecX}_{n,k}(t-\tk),t-\tau\right) \right]_{\tau\,=\,\tkk}^{\tau\,=\,\tk} \, , \label{eq:model5}
\end{equation}
where $\smash{\tilde{\vecX}_{n,k}}$ is the solution to
\begin{equation}
	\frac{\mathrm{d}\tilde{\vecX}_{n,k}(s)}{\mathrm{d}s} = \tilde{\vecu}_{n,\vfrac}(\tilde{\vecX}_{n,k}(s),\tk+s) \, , \label{eq:sourceadvection}
\end{equation}
subject to the initial condition $\smash{\tilde{\vecX}_{n,k}(0) = \vecX_n(\tk)}$.\medskip

\begin{figure}\centering
\subfloat[The $n^\mathrm{th}$ particle follows a trajectory $\smash{\vecX_n(t)}$ from $0$ to $t$, while feeding the force $\smash{-\vecF_n(t)}$ back to the fluid]{\includegraphics{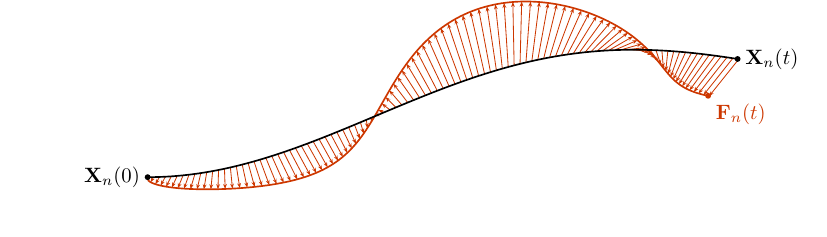}}\\
\subfloat[The time interval $\smash{[0,t]}$ is split into ${K=9}$ intervals $\smash{{[\tk,\tkk]}}$ with $\smash{{t^{_{(1)}} = 0, t^{_{(10)}} = t}}$. The discrete particle positions and feedback forces at the times $\tk$ are identified as $\smash{\vecX_n^{_{(k)}} = \vecX_n(\tk)}$ and $\smash{-\vecF_n^{_{(k)}} = -\vecF_n(\tk)}$.]{\includegraphics{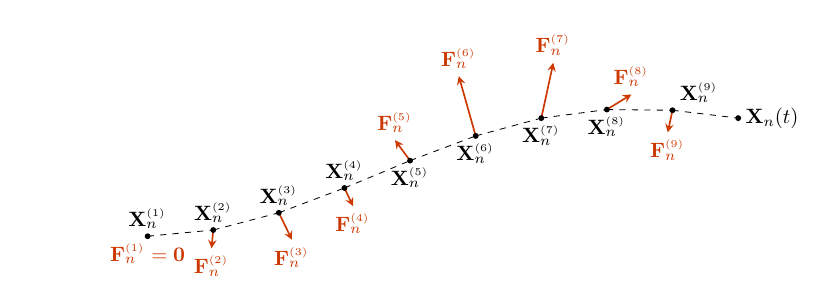}}\\
  \subfloat[In accordance with the model proposed in Eq.~\eqref{eq:model5}, the center of each transient Stokeslet contribution needs to be transported by the undisturbed flow, from its time of injection onwards. These advected source locations, solution to Eq.~\eqref{eq:sourceadvection}, are identified as $\smash{\tilde{\vecX}_n^{_{(k)}} = \tilde{\vecX}_{n,k}(t-\tk)}$.]{\includegraphics{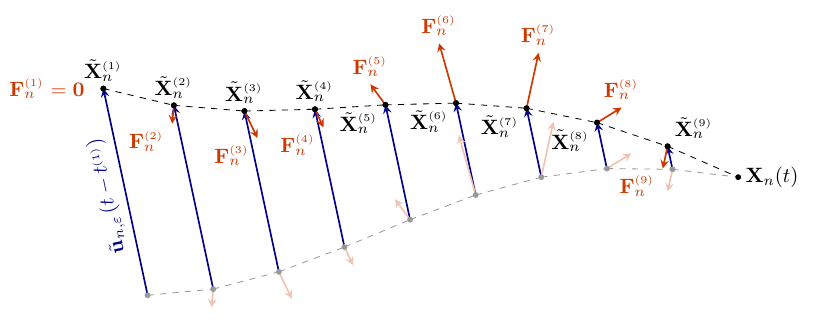}}
  \caption{Illustration of the time discretization of a particle's trajectory, and of the transport of the Stokeslet contributions by the undisturbed flow.} \label{fig:illustrationmodel}
\end{figure}

In practice, this means that:
\begin{itemize}
	\item The velocity disturbance induced by a particle is a function of all previous discrete instances of the feedback force associated with that particle (\ie, it is function of $\vecF_n(\tk), k \in \{1,\ldots,K\}$).
	\item Each of the discrete force instances $\smash{\vecF_n(\tk)}$ contributes to $\smash{{\vecu}^\prime_{n,\vfrac}}$ in the form of a regularized transient Stokeslet ``active'' between $\tk$ and $\tkk$, obtained via the multiplication of $\smash{\vecF_n(\tk)}$ with the tensorial Green's function operator $\smash{\integral{\vecG}_\K}$.
	\item Each of these regularized transient Stokeslet contributions originates from a source-location $\smash{\tilde{\vecX}_{n,k}}$ advected with the undisturbed flow from $\tk$ onwards, where $\tk$ is the time of introduction of the discrete feedback force.
\end{itemize}
The discrete setup of this model is illustrated in Figure~\ref{fig:illustrationmodel}. Note that estimating the undisturbed velocity at each source-location $\smash{\tilde{\vecX}_{n,k}}$ for its advection with Eq.~\eqref{eq:sourceadvection} is tedious, complex to implement, and computationally expensive. Instead, in practice, we advect these sources with the filtered velocity, $\smash{\bar{\vecu}_{\vfrac}}$, rather than the undisturbed velocity, $\smash{\tilde{\vecu}_{n,\vfrac}}$. We have found this to have little effect on the accuracy of the model, while significantly reducing its complexity and computational cost. 

\subsection{Approximating the Laplacian of the particles' self-induced velocity disturbance}
\label{sec:greenlap}
Applying the Laplacian operator to the expression in Eq.~\eqref{eq:green2}, it is straightforward to obtain Green's function for the Laplacian of the transient Stokeslet, also referred to as the transient potential dipole,
\begin{equation}
	\vecpd(\vecx,t) = \frac{1}{\mu} \left(\identity{\dd}_1(\|\vecx\|,t) + \left(\vecx \otimes \vecx\right) \, {\dd}_2(\|\vecx\|,t) \right) \, ,
\end{equation}
where
\begin{align}
	{\dd}_1(r,t) & = \dfrac{1}{4\pi r^3} \left[\dfrac{\left(r^2-4\nu t\right)r^3}{8 \nu^2 t^3\sqrt{\pi \nu t}} \exp\left( -\frac{r^2}{4\nu t} \right) \right]\, , \\
	{\dd}_2(r,t) & = \dfrac{-3}{4\pi r^5} \left[\dfrac{r^5}{24 \nu^2 t^3 \sqrt{\pi \nu t}} \exp\left( -\frac{r^2}{4\nu t} \right) \right] \, .
\end{align}
The functions $\smash{{\dd}_1}$ and $\smash{{\dd}_2}$ are singular both in space, at $\smash{r = \|\vecx\| = {0}}$, and in time, at $t=0$.
Following the exact same steps as in Section~\ref{sec:modeldist}, we can derive an approximation of the Laplacian of the $\vfrac$-weighted velocity disturbance, which reads as
\begin{equation}
	\nabla^2{\vecu}^\prime_{n,\vfrac} (\vecx,t) = \sum_{k=1}^{K} \vecF_n\left(\tk\right) \cdot \left[  \integral{\vecPD}_\K\left(\vecx-\tilde{\vecX}_{n,k}(t-\tk),t-\tau\right) \right]_{\tau\,=\,\tkk}^{\tau\,=\,\tk} \, , \label{eq:model6}
\end{equation}
where
\begin{equation}
	\integral{\vecPD}_\K = \K \Conv \integral{\vecPD} \, ,
\end{equation}
and
\begin{equation}
	\integral{\vecPD}(\vecx,t) = \int_{0}^{t} {\vecpd}(\vecx,s) \, \mathrm{d}s = \frac{1}{\mu} \left(\identity\integral{\D}_1(\|\vecx\|,t) + \left(\vecx \otimes \vecx\right) \, \integral{\D}_2(\|\vecx\|,t) \right) \, .
\end{equation}
In the previous expression,
\begin{align}
	\integral{\D}_1(r,t) & =\int_{0}^{t} {\dd}_1(r,s) \, \mathrm{d}s=  \dfrac{1}{4\pi r^3} \left[1 + \dfrac{r\left(r^2 + 2\nu t\right)}{2 \nu t \sqrt{\pi \nu t} } \exp\left( -\frac{r^2}{4\nu t} \right) - \erf \left( \frac{r}{\sqrt{4\nu t}}\right)\right] \, ,\\
	\integral{\D}_2(r,t) & = \int_{0}^{t} {\dd}_2(r,s) \, \mathrm{d}s= \dfrac{-3}{4\pi r^5} \left[1 + \dfrac{r\left(r^2 + 6\nu t\right)}{6 \nu t \sqrt{\pi \nu t} } \exp\left( -\frac{r^2}{4\nu t} \right) - \erf \left( \frac{r}{\sqrt{4\nu t}}\right) \right]\, .
\end{align}

Alternatively, an approximation of $\smash{\nabla^2{\vecu}^\prime_{n,\vfrac}}$ could also be obtained by directly applying a discrete Laplacian operator to the vector field $\smash{{\vecu}^\prime_{n,\vfrac}}$ given by Eq.~\eqref{eq:model5}. 

\section{Special cases}
\label{sec:special_cases}
In this section are discussed special cases of isolated particles for which the model proposed in Section~\ref{sec:modeldist} recovers well-known analytical solutions found in the literature.

\subsection{Fixed source in quiescent flow} \label{sec:specialcase1} 
Let us consider the case in which:
\begin{itemize}
	\item The Reynolds number associated with the evolution of the $n^\mathrm{th}$ particle and its transfer of momentum is very small, \ie the particle's effect on the flow can be considered as that of a fixed momentum source in quiescent fluid;
	\item The particle feeds a constant force $\smash{-\vecF_n}$ back to the fluid.
\end{itemize}
In such a case, the linearized equations governing the particle-induced flow disturbance, in the frame of reference attached to the particle, read as
\begin{align}
	\nabla \cdot {\vecv}^\prime_{n,\vfrac} & = 0 \, , \label{eq:lowREdistcont} \\
	\rho \frac{\partial {\vecv}^\prime_{n,\vfrac}}{\partial t} & = -\nabla {p}^{\prime}_{n,\vfrac} + \mu \nabla^2 {\vecv}^\prime_{n,\vfrac} - \K\left(\left\|\vecy\right\|\right) \vecF_n \, . \label{eq:lowREdistmom}
\end{align}
The particle-induced, $\vfrac$-weighted velocity disturbance is thus given by
\begin{equation}
	{\vecv}^\prime_{n,\vfrac} (\vecy,t) = -\vecF_n \cdot \integral{\vecG}_\K(\vecy,t) \, ,
\end{equation}
which is simply the expression of the persistent, transient regularized Stokeslet centered at the location of the $n^\mathrm{th}$ particle. As $t$ tends to infinity, we recover the expression of the steady regularized Stokeslet, as used for instance in \cite{Balachandar2019,Evrard2020a}.

\subsection{Fixed source in steady uniform flow} \label{sec:specialcase2}
Let us now consider the case in which:
\begin{itemize}
	\item The particle's effect on the flow can be considered as that of a fixed momentum source subject to uniform steady flow;
	\item The particle feeds a constant force $\smash{-\vecF_n}$ back to the fluid, which is parallel to the undisturbed flow.
\end{itemize}
This corresponds, for instance, to the case of an isolated particle falling at terminal velocity in quiescent fluid. In such a case, the flow reaches a steady state for which the linearized equations governing the particle-induced disturbance, in the frame of reference attached to the particle, read as
\begin{align}
	\nabla \cdot {\vecv}^\prime_{n,\vfrac} & = 0 \, , \label{eq:oseendistcont} \\
	\rho \tilde{\vecv}_{n,\vfrac} \cdot \nabla {\vecv}^\prime_{n,\vfrac} & = -\nabla {p}^{\prime}_{n,\vfrac} + \mu \nabla^2 {\vecv}^\prime_{n,\vfrac} - \K\left(\left\|\vecy\right\|\right) \vecF_n \, . \label{eq:oseendistcmom}
\end{align}
The particle-induced $\vfrac$-weighted velocity disturbance is thus given by
\begin{equation}
	{\vecv}^\prime_{n,\vfrac} (\vecy,t) = -\vecF_n \cdot {\vecO}^{\infty}_\K(\vecy,\tilde{\vecv}_{n,\vfrac}) \, , \label{eq:regoseenlet}
\end{equation}
where $\smash{\vecO^{\infty}_\K}$ is the regularized Oseenlet operator obtained from the spatial convolution of $\K$ with the singular Oseenlet operator, $\smash{\vecO^{\infty}}$, as done in \cite{Balachandar2019,Evrard2020a}. Fundamentally, the Oseenlet operator can be obtained by time convolution of the transient Stokeslet advected at constant velocity in the direction of the uniform flow \cite{Chan2000}. Therefore, in the limit of infinitesimal steps $\smash{\tkk-\tk}$, the model provided in Eq.~\eqref{eq:model4} converges towards the steady regularized Oseenlet as given in Eq.~\eqref{eq:regoseenlet}.

\section{Numerics}
\label{sec:numerics}
In this section, we discuss the implementation of the model proposed in Eq.~\eqref{eq:model5} and study its convergence.

\subsection{Determining the amount of previous instances needing to be stored} \label{sec:timelimit}
Storing all previous forcing instances considered in the sum of Eq.~\eqref{eq:model5} evidently generates untractable computational costs when simulating particle-laden flows containing a significant amount of particles. Owing to the asymptotic behavior of the tensorial operator $\smash{\integral{\vecG}_\K}$ in time, however, it is clear that the most ``recent'' forcing instances are those mainly contributing to the particle-induced velocity disturbance, while the contributions of the ``old'' forcing instances become less and less significant as time increases.\medskip

\begin{figure}\centering
	\includegraphics{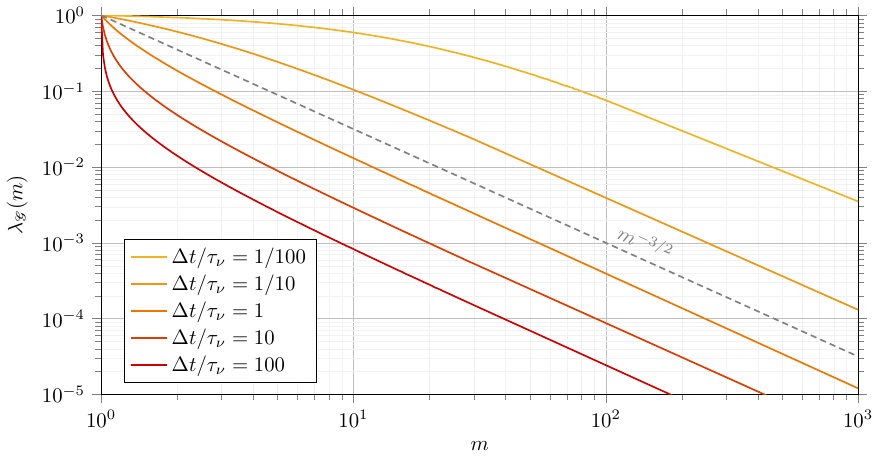}
	\caption{Importance of the $m^\text{th}$ previous time instance relative to the most recent time instance, in the discrete temporal convolution sum of Eq.~\eqref{eq:model5}.}
	\label{fig:ratiotophat}
\end{figure}

For any monotonically decreasing radial kernel $\K$, the operator $\smash{\integral{\vecG}_\K}$ will typically generate maximum velocity contributions at $\smash{\vecx=\boldsymbol{0}}$, where
\begin{equation}
	\integral{\vecG}_\K (\boldsymbol{0},t) = \int_{\mathbb{R}^3} \K(\|\vecy\|) \integral{\vecG}(\vecy,t) \, \mathrm{d}{\vecy} = \identity\left[\frac{4\pi}{\mu} \int_0^\infty \K(r)\left(r^2\integral{\Hh}_1(r,t) + \frac{r^4}{3}\integral{\Hh}_2(r,t)\right)  \, \mathrm{d}r\right]\, . \label{eq:regstokeslet0}
\end{equation}
The scalar integral in Eq.~\eqref{eq:regstokeslet0},
\begin{equation}
	\mathscr{S}_{\K_0} (t) = \frac{4\pi}{\mu}\int_0^\infty \K(r)\left(r^2\integral{\Hh}_1(r,t) + \frac{r^4}{3}\integral{\Hh}_2(r,t)\right)  \, \mathrm{d}r \, , \label{eq:intoriginstokeslet}
\end{equation}
can generally be derived into a closed-form expression (see examples in \ref{apdx:manualconv}), and the ratio
\begin{equation}
	\lambda_\K (m) = \dfrac{\mathscr{S}_{\K_0} (m\Delta t) - \mathscr{S}_{\K_0} ((m-1)\Delta t)}{\mathscr{S}_{\K_0} (\Delta t)} \, 
\end{equation}
then approximates the importance of the $m^\text{th}$ previous time instance relative to the most recent time instance in the discrete temporal convolution sum of Eq.~\eqref{eq:model5}. It can be shown that as $m$ goes to infinity, the ratio $\smash{\lambda_\K (m)}$ tends to zero as
\begin{equation}
	\lambda_\K (m)
  \underset{\tiny\text{$m\! \to \!\infty$}}{\propto}
  {m^{-3/2}} \ .
\end{equation}
For instance if $\K$ is chosen as the gaussian filter $\mathscr{G}$ of standard deviation $\sigma$, as defined in Eq.~\eqref{eq:kernelgaussian}, then
\begin{equation}
	\lambda_\mathscr{G} (m) = \dfrac{\left(2 (m-1) \Delta t / \tau_\nu + \beta^2\right)^{-1/2} - \left(2 m\Delta t/ \tau_\nu + \beta^2\right)^{-1/2}}{\beta^{-1} - \left(2 \Delta t / \tau_\nu + \beta^2\right)^{-1/2}} \, ,
\end{equation}
where $\tau_\nu$ is the viscous time-scale associated with the filtering kernel defined as
\begin{equation}
	\tau_\nu = {\ell^2}/{\nu} \, ,  \label{eq:defviscoustau}
\end{equation}
$\beta = {\sigma}/{\ell}$, 
and $\ell$ is the filter length-scale defined by Eq.~\eqref{eq:defell}. The ratio $\smash{\lambda_\mathscr{G}}$ is plotted for different values of $\smash{\Delta t/\tau_\nu}$ in Figure~\ref{fig:ratiotophat}. From this figure, it is clear that the smaller $\smash{\Delta t/\tau_\nu}$ is, the more previous time instances must be stored to accurately estimate the self-induced velocity disturbance according to Eq.~\eqref{eq:model5}. When $\smash{\Delta t = \tau_\nu}$, the $20^\text{th}$ previous forcing instance in Eq.~\eqref{eq:model5} accounts for approximately less than $1\%$ of the most recent forcing instance. When $\smash{\Delta t = 100 \tau_\nu}$, the $3^\text{rd}$ previous forcing instance in Eq.~\eqref{eq:model5} already accounts for less than $1\%$ of the most recent forcing instance.

\subsection{Computing the regularized transient Stokeslet/potential dipole tensorial operators}
\label{sec:discreteconv}
\begin{figure}[!b]\centering
  \begin{tabular}{>{\centering\arraybackslash} m{0.28\linewidth} >{\centering\arraybackslash} m{0.01\linewidth} >{\centering\arraybackslash} m{0.28\linewidth} >{\centering\arraybackslash} m{0.01\linewidth} >{\centering\arraybackslash} m{0.28\linewidth}}
  \includegraphics{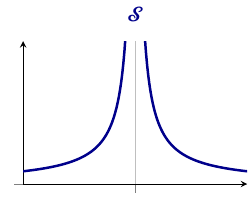}
  & {\Large $\Conv$} &
  \includegraphics{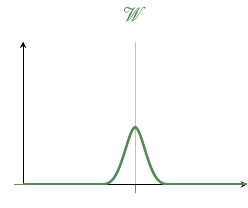}
  &  {\Large $=$} &
  \includegraphics{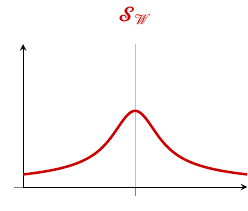}
  \end{tabular}\vspace{-\baselineskip}
  \begin{center}
    \scriptsize (a) The regularized Stokeslet operator is the convolution of the singular Stokeslet operator with\\ the regularization/filtering kernel (here chosen as the Wendland kernel $\mathscr{W}$).  
  \end{center}
    \begin{tabular}{>{\centering\arraybackslash} m{0.28\linewidth} >{\centering\arraybackslash} m{0.01\linewidth} >{\centering\arraybackslash} m{0.28\linewidth} >{\centering\arraybackslash} m{0.01\linewidth} >{\centering\arraybackslash} m{0.28\linewidth}}
      \includegraphics{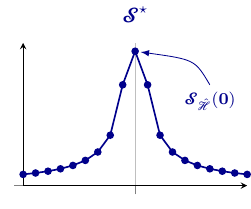}
      & {\Large $\Conv$} &
      \includegraphics{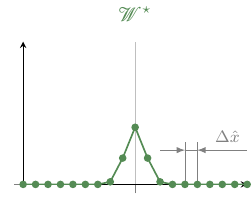}
      &  {\Large $=$} &
      \includegraphics{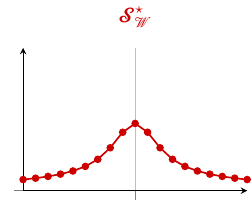}
      \end{tabular}\vspace{-\baselineskip}
      \begin{center}
        \scriptsize (b) The \underline{discrete} regularized Stokeslet operator is computed as the inverse DFT of the product\\ of the DFTs of the discrete singular Stokeslet operator with the discrete regularization/filtering kernel.
      \end{center}
      \caption{Schematic of the discrete convolution of the singular Stokeslet operator with the regularization kernel, highlighting the discrete sampling of the functions.}
  \label{fig:mapscreation}
\end{figure}

\begin{figure}\raggedright
 \includegraphics{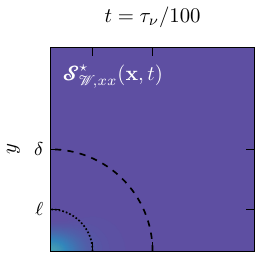}
 \includegraphics{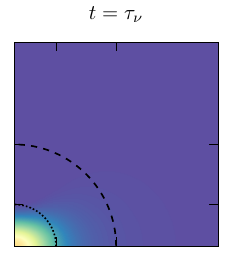}
 \includegraphics{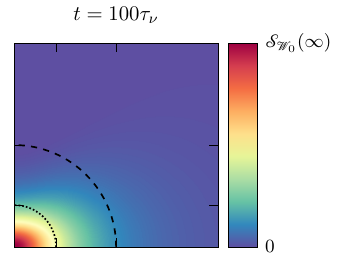}\\
 \includegraphics{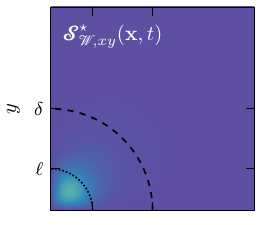}
 \includegraphics{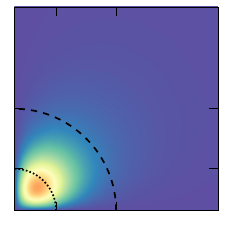}
 \includegraphics{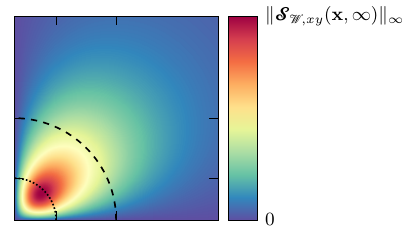}\\
 \includegraphics{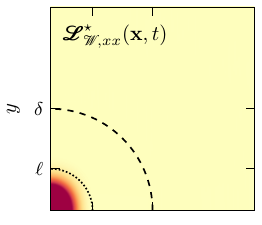}
 \includegraphics{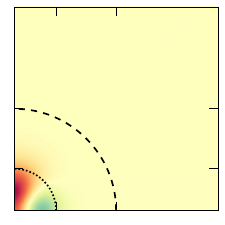}
 \includegraphics{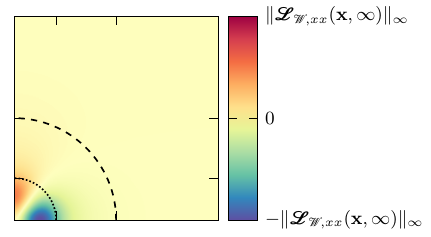}\\
 \includegraphics{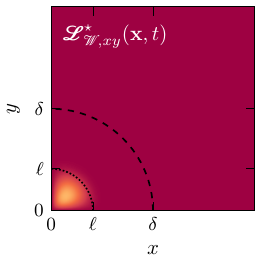}
 \includegraphics{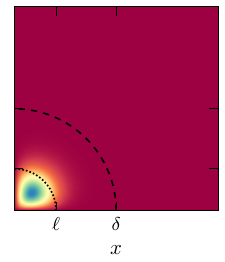}
 \includegraphics{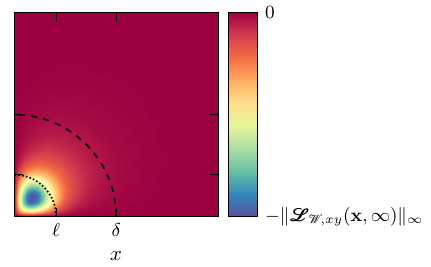}\\
\caption{Example of discrete maps for the regularized transient Stokeslet and potential dipole tensors at three different times $t \in \{\tau_\nu/100,\tau_\nu,100\tau_\nu\}$, and with $\K$ chosen as the Wendland polynomial kernel $\mathscr{W}$, defined in Eq.\eqref{eq:defwendland}, compactly supported on the ball of radius $\delta$. The length-scale $\ell$ is defined by Eq.~\eqref{eq:defell}. The time-scale $\tau_\nu$ is defined by Eq.~\eqref{eq:defviscoustau}.}
\label{fig:mapsexamples}
\end{figure}

Deriving analytical expressions for the regularized transient Stokeslet and potential dipole tensorial operators, $\smash{\integral{\vecG}_\K}$ and $\smash{\integral{\vecPD}_\K}$, is tedious or often even simply impossible. Their discrete computation, on the other hand, presents several advantages:
\begin{itemize}
	\item Its implementation is relatively straightforward;
  \item It enables an arbitrary choice of filtering kernel $\K$;
	\item It naturally adapts to the finite spatial and temporal resolutions of the Eulerian variable fields (and therefore of the particle-induced velocity disturbance generated by the numerical solution of the volume-filtered governing equations), if the same spatial resolution is used for the discrete convolution of the singular tensorial operators with the filtering kernel $\K$, and the same temporal resolution is used for their sampling over time;
	\item The discrete maps of the regularized tensorial operators can be computed once and for all in a pre-simulation step and stored in memory, resulting in a negligible cost overall;
	\item Using the symmetries of the filtering kernel and singular tensorial operators, the computational memory footprint of these solutions can be minimised.
\end{itemize}
We thus choose to discretely map, both in space and time, the regularized transient Stokeslet and potential dipole tensorial operators, $\smash{\integral{\vecG}_\K}$ and $\smash{\integral{\vecPD}_\K}$. To that end, the singular tensorial operators, $\smash{\integral{\vecG}}$ and $\smash{\integral{\vecPD}}$, and the filtering kernel, $\K$, are spatially sampled on a grid of resolution $\Delta \hat{x}$ before being spatially convoluted. The resulting discrete maps of the regularized transient Stokeslet and potential dipole tensorial operators are generated at discrete times that are either sampled uniformly or logarithmically in time. If $\Delta \hat{x}$ is chosen smaller than the Eulerian grid resolution $\Delta {x}$, the regularized operators $\smash{\integral{\vecG}_\K}$ and $\smash{\integral{\vecPD}_\K}$ are subjected to an additional spatial convolution with the top-hat kernel ${\mathscr{H}}$ of radius $\smash{\sqrt[3]{3/4\pi}\Delta {x}}$, so as to match the regularization imposed by our second-order finite-volume discretization. As mentioned in Section~\ref{sec:model}, the singular tensorial operators $\smash{\integral{\vecG}}$ and $\smash{\integral{\vecPD}}$ are regular in time, but singular in space at $\smash{\vecx=\boldsymbol{0}}$. With the aim to sample these two operators with (close to) second-order spatial accuracy, we employ the following strategy to calculate their discrete values at the sampling grid-points $\vecx_i$:
\begin{itemize}
  \item If $\left\|\vecx_i\right\|<\Delta \hat{x} /2$,\vspace{-1mm}
  \begin{equation}
    {\integral{\vecG}}^\star_{i}(t) = \integral{\vecG}_{\hat{\mathscr{H}}} (\boldsymbol{0},t) \, ,
  \end{equation}
  where $\smash{\hat{\mathscr{H}}}$ is the top-hat filter compact on the ball of radius $\smash{\sqrt[3]{3/4\pi}\Delta \hat{x}}$. The quantity $\integral{\vecG}_\mathscr{H} (\boldsymbol{0},t)$ is derived in \ref{apdx:manualconvtophat}.
  \item Otherwise,\vspace{-1mm}
  \begin{equation}
    {\integral{\vecG}}^\star_{i}(t) = \integral{\vecG} (\vecx_i,t) \, .
  \end{equation}
\end{itemize}
A similar strategy is employed to spatially sample $\smash{\integral{\vecPD}}$. The filtering kernel $\K$ is spatially sampled in a similar manner as for the computation of $\smash{\vfrac}$ or $\smash{\vecf}$ on the Eulerian grid. This discrete sampling and convolution process is illustrated in Figure~\ref{fig:mapscreation}.
Owing to the radiality of the filtering kernel $\K$ and the symmetries of $\smash{\integral{\vecG}}$ and $\smash{\integral{\vecPD}}$, it is sufficient to map one diagonal and one off-diagonal component of each tensor $\smash{\integral{\vecG}_\K}$ and $\smash{\integral{\vecPD}_\K}$ in the top-right quadrant of the two-dimensional real space. Examples of the resulting two-dimensional maps are given in Figure~\ref{fig:mapsexamples}.
Although the discrete convolutions of $\integral{\vecG}$ and $\integral{\vecPD}$ with $\K$ can be computed ``naturally'', \ie by calculating the convolution integrals with some numerical quadrature rule, the use of discrete Fourier transforms (DFT) can hugely speed up the process of generating the maps\footnote{The convolution theorem specifies that the convolution of two functions is obtained from the inverse Fourier transform of the product of their respective Fourier transforms; $\phi \Conv \psi = \matholdcal{F}^{-1} \left( \matholdcal{F}(\phi) \cdot \matholdcal{F}(\psi)\right)$.}.\medskip

In practice, after having generated the discrete maps, the following cases arise when interpolation $\smash{\integral{\vecG}_\K(\vecx,t)}$ or $\smash{\integral{\vecPD}_\K(\vecx,t)}$ from them:
\begin{itemize}
  \item If $\vecx$ and $t$ are located within the spatial and temporal scopes of the maps, then we use quadri-linear interpolation from the maps;
  \item If $\vecx$ is outside the spatial scope of the maps, then we use the approximations $\smash{\integral{\vecG}_\K(\vecx,t) \simeq \integral{\vecG}(\vecx,t)}$ and $\smash{\integral{\vecPD}_\K(\vecx,t) \simeq \integral{\vecPD}(\vecx,t)}$, since far away from the singularity, the regularized tensorial operators converge towards their singular equivalents (see, e.g., Figure~\ref{fig:compsingreg});
  \item If $t$ is outside the temporal scope of the maps and $\vecx$ is within the spatial scope of the maps, we use the steady regularized tensorial operator (see, e.g., Eq.~\eqref{eq:steadywendland} for an expression of the steady tensorial operator regularized by the Wendland kernel). 
\end{itemize}

\subsection{Spatial convergence}
\begin{figure}\centering
  \includegraphics{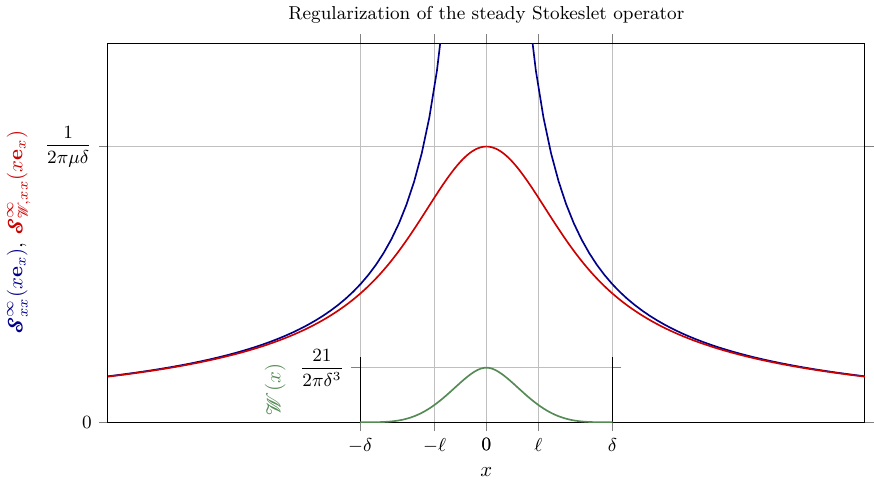}
\caption{Diagonal coefficient of the singular and regularized Stokeslet operators, along the $x$-axis. The regularization kernel is the Wendland kernel, $\smash{\mathscr{W}}$, defined in Eq.~\eqref{eq:defwendland}. The length-scale $\ell$ is defined by Eq.~\eqref{eq:defell}.}
\label{fig:compsingreg}
\end{figure}

In this section, our aim is to assess the spatial convergence of the proposed model and, more generally, its performance at finite spatial resolution. To that end, we consider the spatial convolution of the singular, steady Stokeslet operator,
\begin{equation}
  \vecG^{\infty}(\vecx) = \dfrac{1}{8\pi\mu\|\vecx\|} \left(\identity + \dfrac{\vecx\otimes\vecx}{\|\vecx\|^2}\right) \, ,
\end{equation}
with the Wendland kernel defined in Eq.~\eqref{eq:defwendland}. The corresponding regularized operator can be derived analytically \cite{Cortez2001,Evrard2020a}, and reads as
\begin{equation}
  \vecG^{\infty}_{\mathscr{W}}(\vecx) = \left[ \vecG^{\infty} \Conv \mathscr{W} \right](\vecx) = \dfrac{1}{8\pi \mu} \left(\identity \integral{\Hh}^{\infty}_{1,\mathscr{W}}(\|\vecx\|) + \left(\vecx\otimes\vecx\right) \integral{\Hh}^{\infty}_{2,\mathscr{W}}(\|\vecx\|)\right) \, , \label{eq:steadywendland}
\end{equation}
where
\begin{align}
  \integral{\Hh}^{\infty}_{1,\mathscr{W}}(r) & = \left\{ \begin{array}{ll}
    \dfrac{-81r^7 + 400r^6\delta - 735r^5\delta^2 + 540r^4\delta^3 - 168r^2\delta^5 + 60\delta^7}{15\delta^8}, & \quad \text{if } r < \delta \\
    \dfrac{1}{r} + \dfrac{\delta^2}{15r^3}, & \quad \text{if } r \ge \delta
  \end{array}\right. \, ,\\
  \integral{\Hh}^{\infty}_{2,\mathscr{W}}(r) & = \left\{ \begin{array}{ll}
    \dfrac{21r^5 - 100r^4\delta + 175r^3\delta^2 - 120r^2\delta^3 + 28\delta^5}{5\delta^8}, & \quad \text{if } r < \delta \\
    \dfrac{1}{r^3} -\dfrac{\delta^2}{5r^5}, & \quad \text{if } r \ge \delta
  \end{array}\right. \, .
\end{align}
The first diagonal component of these singular and regularized operators, along the $x$-axis, is shown in Figure~\ref{fig:compsingreg}. Note that we recover the classical result that, at distances much larger than the length-scale associated with the convolution kernel, the regularized Stokeslet operator converges towards the singular Stokeslet, 
\begin{equation}
  \lim\limits_{\|\vecx\| \to\infty} \left(\vecG^{\infty}_{\mathscr{W}}(\vecx) - \vecG^{\infty}(\vecx)\right) = \tensor{0} \, .
\end{equation}
As described in Section~\ref{sec:discreteconv}, the proposed discrete execution of the spatial convolution between $\smash{\mathscr{W}}$ and $\smash{\vecG^\infty}$ requires their sampling on a grid of uniform spacing $\Delta \hat{x}$. The discrete samples of $\smash{\mathscr{W}}$ are computed with the adaptive quadrature rule presented in \citep{Evrard2019}. The discrete samples of $\smash{\vecG^\infty}$ are computed as described in Section~\ref{sec:discreteconv}, in order to treat the singularity at $\smash{\vecx=\boldsymbol{0}}$. This means that the singular steady Stokeslet is sampled at $\vecx_i$, the location of the $i^\mathrm{th}$ sampling point, as follows:
\begin{equation}
  \integral{\vecG}^{\star,\infty}_{i} = \left\{ \begin{array}{ll}
  \identity\left(4\pi\mu\alpha\Delta \hat{x}\right)^{-1}  & \quad \text{if } \left\|\vecx_i\right\|<\Delta \hat{x} /2 \\ & \\
  \integral{\vecG}^{\infty} (\vecx_i)& \quad \text{otherwise}
  \end{array}\right. \, ,
\end{equation}
with $\smash{\alpha=\sqrt[3]{3/(4\pi)}}$.
Once $\smash{\mathscr{W}}$ and $\smash{\vecG^\infty}$ have been discretely sampled into $\smash{\mathscr{W}}^\star$ and $\smash{\vecG^{\star,\infty}}$, the corresponding (discrete) regularized operator is obtained by computing the inverse DFT of the product of their DFTs. \medskip

\begin{figure} \centering
\subfloat[Spatial convolution error\label{fig:spatialconv}]{
  \includegraphics{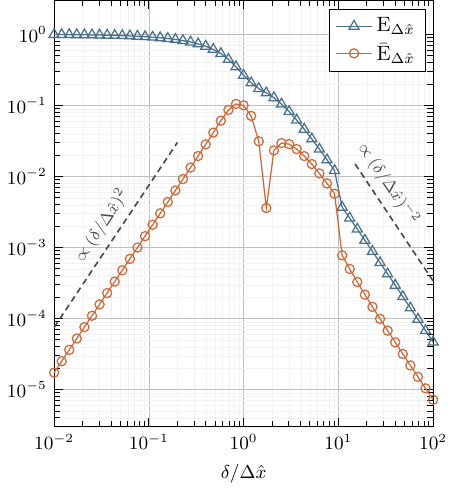}
}\subfloat[Temporal convolution error\label{fig:temporalconv}]{
  \includegraphics{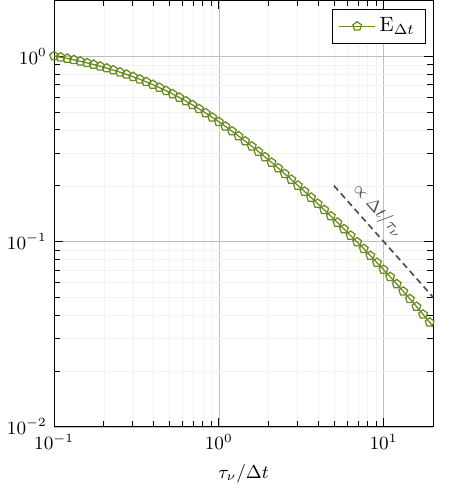}
}
\caption{In both graphs, the Wendland kernel defined in Eq.~\eqref{eq:defwendland} is used as regularization kernel: (a) Relative errors between the discrete and analytical regularized steady Stokeslet operators, computed at the origin. The error is shown as a function of the ratio between $\smash{\delta}$, the radius of the Wendland kernel's support, and $\smash{\Delta x}$, the spacing of the sampling grid for the discrete convolution; (b) Relative error between the discrete and analytical regularized Oseenlet operators, computed at the origin. The error is shown as a function of the ratio between the viscous time-scale of the kernel, $\smash{\tau_\nu}$, and $\smash{\Delta t}$, the timestep used to discretize the temporal convolution integral.}
\end{figure}

In order to determine the order of spatial convergence of the proposed model, the relative error between the discrete regularized operator and its analytical counterpart is computed at $\vecx_0 = \boldsymbol{0}$ (where both operators are diagonal, by construction). We consider two errors, defined as
\begin{align}
  \mathrm{E}_{\Delta \hat{x}} & = \dfrac{\left|\mathrm{tr}(\vecG^{\star,\infty}_{\mathscr{W}\!,0}) - \mathrm{tr}(\vecG^{\infty}_{\mathscr{W}}(\boldsymbol{0}))\right|}{\left|\mathrm{tr}(\vecG^{\infty}_{\mathscr{W}}(\boldsymbol{0}))\right|} \, , \\
  \bar{\mathrm{E}}_{\Delta \hat{x}} & = \dfrac{\left|\mathrm{tr}(\vecG^{\star,\infty}_{\mathscr{W}\!,0}) - \mathrm{tr}(\vecG^{\infty}_{\mathscr{W},\hat{\mathscr{H}}}(\boldsymbol{0}))\right|}{\left|\mathrm{tr}(\vecG^{\infty}_{\mathscr{W},\hat{\mathscr{H}}}(\boldsymbol{0}))\right|} \, .
\end{align}
The former error, $\smash{\mathrm{E}_{\Delta \hat{x}}}$, is the relative difference between the discrete regularized Stokeslet in the cell centered at $\vecx_0 = \boldsymbol{0}$ and the exact corresponding regularized Stokeslet evaluated at $\vecx_0$. The latter error, $\smash{\bar{\mathrm{E}}_{\Delta \hat{x}}}$, is the relative difference between the discrete regularized Stokeslet in the cell centered at $\vecx_0$ and an approximation of the average of the exact regularized Stokeslet in the cell centered at $\vecx_0$ (this average is estimated by convolution of the exact regularized Stokeslet with the top-hat kernel of radius $\smash{\sqrt[3]{3/4\pi}\Delta \hat{x}}$). At high resolution ($\smash{\delta/\Delta \hat{x} \gg 1}$), we should expect both errors to exhibit a similar behavior, since the difference between the exact regularized Stokeslet evaluated at $\vecx_0$ and its average in the cell centered at $\vecx_0$ decreases as resolution increases. At low resolution ($\smash{\delta/\Delta \hat{x} \ll 1}$) we should expect both errors to behave differently: The error $\smash{\mathrm{E}_{\Delta \hat{x}}}$ should reach the asymptotic value of $1$, since 
\begin{equation}
  \lim\limits_{\delta/\Delta \hat{x} \to 0} \vecG^{\star,\infty}_{\mathscr{W}\!,0} = \tensor{0} \, ,
\end{equation}
while $\smash{\vecG^{\infty}_{\mathscr{W}}(\boldsymbol{0}) = \identity\left(2\pi \delta\mu\right)^{-1} \ne \tensor{0}}$. The error $\smash{\bar{\mathrm{E}}_{\Delta \hat{x}}}$, on the other hand, should converge towards zero as $\smash{\delta/\Delta \hat{x}}$ decreases. These behaviors are indeed verified by the results plotted in Figure~\ref{fig:spatialconv}.

\subsection{Temporal convergence}
\begin{figure}\centering
  \includegraphics{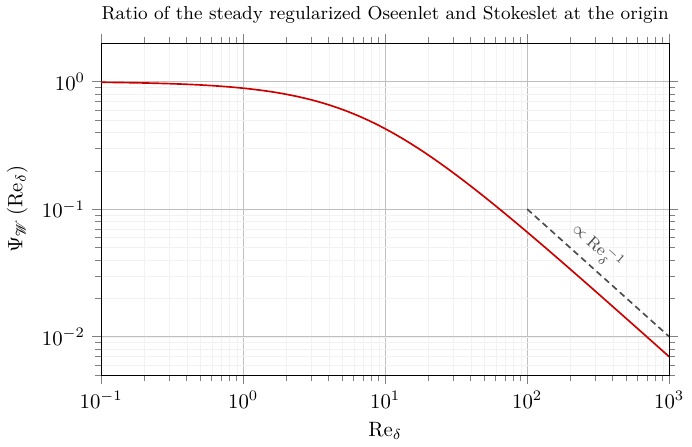}
 \caption{Ratio of the steady regularized Oseenlet and Stokeslet at the origin $\smash{\vecx = 0}$ for the component aligned with the background flow velocity vector $\smash{\tilde{\vecu}}$. At low Reynolds number, this ratio is close to $1$, meaning that the steady regularized Oseenlet and Stokeslet operators are identical. At high Reynolds number, this ratio becomes inversely proportional to the Reynolds number~\cite{Balachandar2019}, meaning that the magnitude of the Oseenlet vanishes compared to the Stokeslet.}
 \label{fig:psioseen}
\end{figure}

In this section, our aim is to assess the temporal convergence of the proposed model. To that end, we consider the spatial convolution of the singular, steady Oseenlet operator \citep{Pozrikidis2011},
\begin{equation}
  \vecO^{\infty} (\vecx) = \dfrac{1}{8\pi\mu}\left(\identity \mathscr{Q}^\infty_1(\vecx) -\nabla\mathscr{Q}^\infty_2(\vecx)\right) \, ,
\end{equation}
where
\begin{align}
  \mathscr{Q}^\infty_1(\vecx) & = \dfrac{2\exp\left( -\xi(\vecx) \right)}{\|\vecx\|}  \, , \\
  \mathscr{Q}^\infty_2(\vecx) & = \dfrac{1-\exp\left( -\xi(\vecx) \right)}{\xi(\vecx)}\left( \dfrac{\vecx}{\|\vecx\|} - \dfrac{\tilde{\vecu}}{\|\tilde{\vecu}\|} \right)  \, , \\
  \xi(\vecx) & = \dfrac{\|\tilde{\vecu}\|\|\vecx\| - \tilde{\vecu}\cdot\vecx}{2\nu} \, , 
\end{align}
with the Wendland kernel defined in Eq.~\eqref{eq:defwendland}. Assuming that $\smash{\tilde{\vecu}}$ is aligned with $\smash{\mathbf{e}_x}$, the first diagonal component of the regularized Oseenlet operator, at $\smash{\vecx = \boldsymbol{0}}$, reads as \cite{Balachandar2019,Evrard2020a}
\begin{equation}
  \vecO^{\infty}_{\mathscr{W}\!,xx} (\boldsymbol{0}) = \dfrac{\Psi_\mathscr{W}\left(\text{Re}_\delta\right)}{2\pi \delta\mu} \, , \label{eq:oseenletanalytical}
\end{equation}
where
\begin{equation}
  \text{Re}_\delta = \dfrac{\delta \|\tilde{\vecu}\|}{\nu} \, ,
\end{equation}
$\delta$ being the radius of the Wendland kernel's support, and $\smash{\Psi_\mathscr{W}}$ is the function given by
\begin{equation}
  \Psi_\mathscr{W}(x) = 
  7\left(x^{-1} 
  - 6x^{-2} 
  + 30x^{-3} 
  - 120x^{-4} 
  + 360x^{-5} 
  - 720x^{-6}
  + 720x^{-7}\left(1-\exp\left(-x\right)\right)\right)\, . \label{eq:psioseen}
\end{equation}
For vanishing Reynolds numbers, 
\begin{equation}
\lim\limits_{\text{Re}_\delta\to0}\Psi_\mathscr{W}\left(\text{Re}_\delta\right) = 1 \, ,
\end{equation}
and one recovers the regularized steady Stokeslet solution derived in \ref{appendix:wendland},
\begin{equation}
  \lim\limits_{\text{Re}_\delta\to0} \vecO^{\infty}_{\mathscr{W}\!,xx} (\boldsymbol{0}) = \vecG^{\infty}_{\mathscr{W}\!,xx} (\boldsymbol{0}) = \frac{1}{2\pi \delta \mu} \, .
\end{equation}
The ratio $\smash{\Psi_\mathscr{W}}$ of the steady Oseenlet and Stokeslet operators regularized by the Wendland kernel, at the location of the origin $\smash{\vecx = \boldsymbol{0}}$, is plotted in Figure~\ref{fig:psioseen} as a function of the Reynolds number $\smash{\text{Re}_\delta}$. Note that the ratio $\Psi_\mathscr{G}$ corresponding to the regularization by the Gaussian kernel $\smash{\mathscr{G}}$, defined in Eq~\eqref{eq:kernelgaussian}, has been derived by \citet{Balachandar2019}, and that its piecewise approximation based on Taylor series expansions that mitigates round-off errors is provided in \cite{Evrard2020a}.\medskip

Importantly for the current temporal convergence analysis and as mentioned in Section~\ref{sec:specialcase2}, it is possible to recover the singular steady Oseenlet operator by calculating the limit, as $\smash{t \to \infty}$, of the time convolution of the singular transient Stokeslet operator, $\smash{\vecg}$, travelling with the velocity $\smash{\tilde{\vecu}}$ \cite{Chan2000}, \ie,
\begin{equation}
  \vecO^{\infty} (\vecx) = \lim\limits_{t\to\infty} \int_0^{t} \vecg(\vecx - (t-\tau) \tilde{\vecu},t-\tau) \, \mathrm{d}\tau \, .
\end{equation}
The regularized Oseenlet operator then reads as
\begin{equation}
  \vecO^{\infty}_{\mathscr{W}} (\vecx) = \lim\limits_{t\to\infty} \int_{\mathbb{R}^3}\int_0^{t} \mathscr{W}(\|\vecx-\vecr\|) \vecg(\vecx - (t-\tau) \tilde{\vecu},t-\tau) \, \mathrm{d}\tau \,\mathrm{d}\vecr \, .
\end{equation}
Applying the same time discretization as in Section~\ref{sec:model}, this regularized operator can be approximated as
\begin{equation}
  \vecO^{\infty}_{\mathscr{W}} (\vecx) \simeq \vecO^{\infty}_{\mathscr{W}\!,\Delta t} (\vecx) =  \sum\limits_{k=1}^{\infty} \left[\vecG_\mathscr{W}(\vecx - \tilde{\vecu}(k-1) \Delta t ,\tau) \right]^{\tau\,=\,k \Delta t}_{\tau\,=\, (k-1)\Delta t} \, , \label{eq:oseenletapprox1}
\end{equation}
where $\smash{\vecG_\mathscr{W}}$ is the time-persistent transient Stokeslet operator regularized by the Wendland kernel. In the context of the volume-filtered modeling of particle-laden flow, this simply corresponds to the (somewhat conceptual) case of a fixed particle, subject to a uniform undisturbed flow $\smash{\tilde{\vecu}}$, and feeding a constant force back to the fluid. In order to study the temporal convergence of the model, we further approximate the previous equation as
\begin{equation}
  \vecO^{\infty}_{\mathscr{W}\!,\Delta t} (\vecx) =  \sum\limits_{k=1}^{K} \left[\vecG_\mathscr{W}(\vecx -\tilde{\vecu}(k-1) \Delta t ,\tau) \right]^{\tau\,=\,k \Delta t}_{\tau\,=\, (k-1)\Delta t} \, , \label{eq:oseenletapprox2}
\end{equation}
where $\smash{K}$ is chosen such that the regularized transient Stokeslet contributions for $k > K$ are negligible.
In Eq.~\eqref{eq:oseenletapprox2}, $\smash{\vecG_\mathscr{W}}$ is interpolated from discrete maps that are generated by the process described in Section~\ref{sec:discreteconv}. We choose $\smash{K}$ such that $\smash{K \Delta t \simeq 1000 \tau_\nu}$ and $\smash{\delta/\Delta x = 100}$, so as to make sure that the error of the model is dominated by the time discretization, instead of the limited number of steps taken or the spatial resolution of the discrete maps. The width of the discrete maps chosen equal to $10 \delta$. Outside this scope, the regularized transient Stokeslet operator, $\smash{\vecG_\mathscr{W}}$, is approximated by its singular (analytical) counterpart, $\smash{\vecG}$, since the difference between these two operators vanishes far away from the singularity. Moreover, we choose $\smash{\text{Re}_\delta = 10}$ in order for the regularized Oseenlet operator, $\smash{\vecO^{\infty}_{\mathscr{W}}}$, to differ greatly from the regularized Stokeslet operator, $\smash{\vecG^{\infty}_{\mathscr{W}}}$. This latter choice of a finite Reynolds number is necessary because, in the limit of a vanishing Reynolds number, Eq.~\eqref{eq:oseenletapprox1} provides an approximation of $\smash{\vecO^{\infty}_{\mathscr{W}}}$ that is independant of the value of $\Delta t$.\medskip

The error of the model due to time discretization is calculated as the relative error between the first diagonal component of the discrete regularized Oseenlet operator, as given by Eq.~\eqref{eq:oseenletapprox2}, and its analytical counterpart, given by Eq.~\eqref{eq:oseenletanalytical},
\begin{equation}
  \mathrm{E}_{\Delta t} = \dfrac{\left|\vecO^{\infty}_{\mathscr{W}\!,\Delta t, xx} (\boldsymbol{0})-\vecO^{\infty}_{\mathscr{W}\!,xx} (\boldsymbol{0})\right|}{\left|\vecO^{\infty}_{\mathscr{W}\!,xx} (\boldsymbol{0})\right|} \, .
\end{equation}
As expected from our use of a left-hand integration rule to estimate the convolution integral in Eq.~\eqref{eq:model3}, the results plotted in Figure~\ref{fig:temporalconv} confirm the first-order accuracy of our discrete time convolution.

\section{Results}
\label{sec:test_cases}
In this section, the model introduced in Section~\ref{sec:model} is used to approximate the particle's self-induced velocity disturbance -- and therefore recover the undisturbed velocity associated with that particle -- for cases of isolated particles with prescribed motion or freely evolving within the fluid.
For ease of implementation, we neglect the unclosed terms $\smash{\boldsymbol{\mathscr{U}}_{\!\mathrm{Re}}}$ and $\smash{\boldsymbol{\mathscr{U}}_{\!\mu}}$, as well as $\smash{\partial \vfrac/\partial t}$ in the volume-filtered governing equations. Thus, they simplify to
\begin{align}
  \nabla \cdot \bar{\vecu}_\vfrac & = 0 \, , \label{eq:masssimple}\\
  \rho \left( \frac{\partial \bar{\vecu}_\vfrac}{\partial t} + \nabla \cdot \left(\bar{\vecu}_\vfrac \otimes \bar{\vecu}_\vfrac \right)\right) & = -\nabla \bar{p}_\vfrac + \mu \nabla^2 \bar{\vecu}_\vfrac + \vecf\, ,\label{eq:momsimple} 
\end{align}
which correspond to the incompressible Navier-Stokes equations for the superficial velocity $\smash{\bar{\vecu}_\vfrac}$ and pressure $\smash{\bar{p}_\vfrac}$, subject to a momentum source $\smash{\vecf}$. Owing to this choice, we can use any conventional incompressible flow solver to test our model, the effect that the particles have on the flow only appearing in the form of the momentum source $\smash{\vecf}$.\medskip

In the following subsections, the second-order finite-volume framework of \citet{Denner2020} is used to solve the set of governing equations~\eqref{eq:masssimple} and \eqref{eq:momsimple}. The Wendland kernel $\smash{\mathscr{W}}$, defined in Eq.~\eqref{eq:defwendland}, is employed for the regularization of the momentum source and is therefore also considered for the production of the discrete regularized transient Stokeslet maps required by the proposed model. Unless specified otherwise, velocity is interpolated at the location of a particle using tri-linear interpolation, and no limit is set for the integer $K$ in Eq.~\eqref{eq:model5}. This means that all previous forcing time instances are considered for approximating the particle's self-induced velocity disturbance at a given time. This choice has been made to showcase the full potential of the model, although having no limit for $K$ would naturally lead to untractable computational cost in the context of real large-scale particle-laden flows. In practice, a limit must be specified for $K$ based on computational cost and/or accuracy considerations (see Sections~\ref{sec:timelimit} and~\ref{sec:history}).

\subsection{Fixed particle in uniform flow} \label{sec:fixedpart}

In a first instance, we consider the case of a fixed particle subject to uniform flow with the constant velocity $\smash{\tilde{\vecu}}$. The particle Reynolds number is defined as
\begin{equation}
  \mathrm{Re}_n = \dfrac{d_n \|\tilde{\vecu}\|}{\nu} \, ,
\end{equation}
where $d_n$ is the diameter of the particle. In each case, the fixed particle feeds back to the fluid a momentum contribution corresponding to the opposite of the steady drag force acting on the particle,
\begin{equation}
  \vecf = -3\pi \mu d_n \tilde{\vecu} f(\mathrm{Re}_n) \mathscr{W}(\|\vecx\|) \, ,
\end{equation}
where $f(\mathrm{Re}_n)$ is the \citet{Schiller1933} empirical correction factor given as
\begin{equation}
  f(\mathrm{Re}_n) = 1 + 0.15 \mathrm{Re}_n^{0.687} \, . \label{eq:schillernaumann}
\end{equation}
The radius of the kernel's support is chosen equal to $2$ particle diameters, i.e., $\smash{\delta = 2 d_n}$. This corresponds to a filter length-scale $\ell \simeq 0.8 d_n$. For each simulation, the computational domain is chosen to be a cubic box with an edge-length equal to or greater than $\smash{L = 100 d_n}$. The center of this domain coincides with the center of the fixed particle, and a constant mesh-spacing $\smash{\Delta x}$ is applied over a distance of at least $3\delta$ on each side of the particle. Outside of this region, the mesh-spacing is stretched so as to avoid unnecessary computational costs.
The following parameter space is considered for choosing $\Delta x$ and the fluid properties:
\begin{itemize}
  \item $d_n / \Delta x \in \{\frac{1}{8}, \frac{1}{4}, \frac{1}{2}, 1, 2, 4\}$
  \item $\mathrm{Re}_n \in \{\frac{1}{100}, \frac{1}{10}, 1, 10, 100\}$
\end{itemize}
Finally, in order to resolve the entire transient evolution of the particle's self-induced flow disturbance, the solver timestep is initially chosen as
\begin{equation}
  \Delta t = \dfrac{\tau_\star}{1000} \, ,
\end{equation}
where
\begin{equation}
  \tau_\star = \dfrac{\ell_\star^2}{\nu} = \tau_\nu \left(\dfrac{\ell}{\ell_\star}\right)^2 \, ,
\end{equation}
and
\begin{equation}
  \ell_\star = \max(\ell,\Delta x \sqrt[3]{3/8\pi} ) \, .
\end{equation}
The modified viscous time-scale $\smash{\tau_\star}$ accounts for the fact that the computational grid acts as a top-hat filter of equivalent radius $\smash{\Delta x \sqrt[3]{3/4\pi}}$, whose length-scale can become larger than $\ell$ at small $\smash{d_n / \Delta x}$ ratio.
At the end of each time iteration, $\smash{\Delta t}$ is increased by a factor $1.1$ unless the CFL limit has been reached.\medskip

\begin{figure} \centering
  \begin{tabular}{l}
    \includegraphics{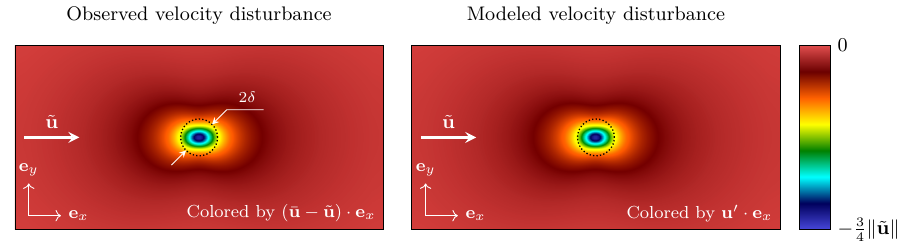}\\
    \includegraphics{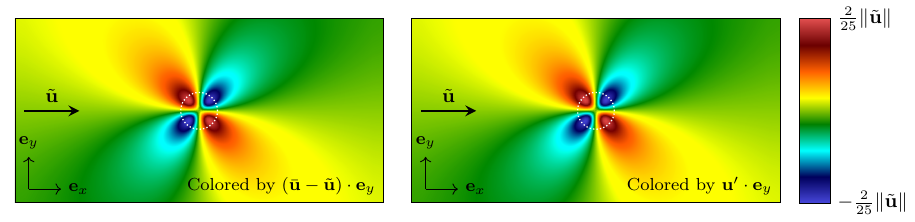}
  \end{tabular}
  \caption{Snapshots of the simulation of a fixed particle subject to uniform flow, once steady state has been reached at $\smash{\mathrm{Re}_n = 0.01}$, with $d_n / \Delta x = 4$ and $\smash{\delta = 2d_n}$. The left column shows the observed velocity disturbance obtained as the solution of the volume-filtered governing equations~\eqref{eq:masssimple} and \eqref{eq:momsimple} from which is subtracted the undisturbed velocity $\smash{\tilde{\vecu}}$. The right column shows its reconstruction obtained with the model proposed in Eq.~\eqref{eq:model5}.} \label{fig:fixedre0p01}
\end{figure}

\begin{figure} \centering
  \begin{tabular}{l}
    \includegraphics{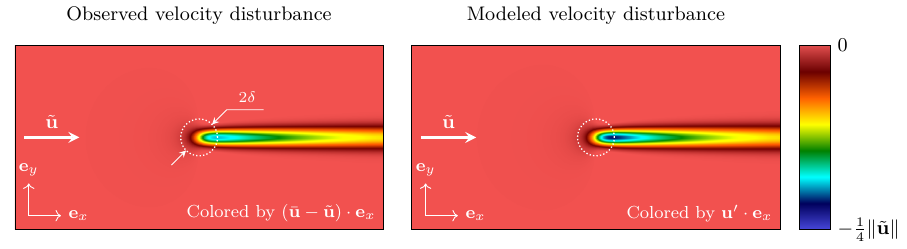}\\
    \includegraphics{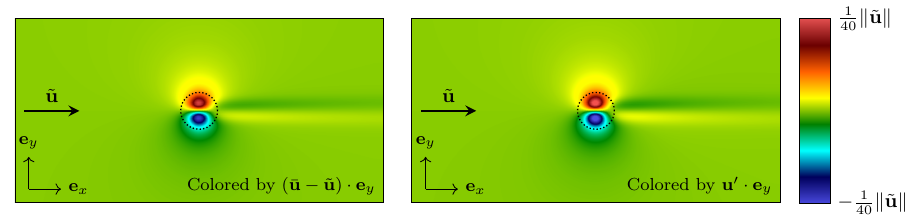}
  \end{tabular}
  \caption{Snapshots of the simulation of a fixed particle subject to uniform flow, once steady state has been reached at $\smash{\mathrm{Re}_n = 100}$, with $d_n / \Delta x = 4$ and $\smash{\delta = 2d_n}$. The left column shows the observed velocity disturbance obtained as the solution of the volume-filtered governing equations~\eqref{eq:masssimple} and \eqref{eq:momsimple} from which is subtracted the undisturbed velocity $\smash{\tilde{\vecu}}$. The right column shows its reconstruction obtained with the model proposed in Eq.~\eqref{eq:model5}.}\label{fig:fixedre100}
\end{figure}

Figures~\ref{fig:fixedre0p01} and \ref{fig:fixedre100} show snapshots of the simulations once a steady state has been reached at $\smash{\mathrm{Re}_n = 0.01}$ and $\smash{\mathrm{Re}_n = 100}$, with $d_n / \Delta x = 4$. The left columns show the observed velocity disturbance induced by the particle, obtained by subtracting the undisturbed velocity $\smash{\tilde{\vecu}}$ from the filtered velocity $\smash{\bar{\vecu}}$ solution to the volume-filtered governing equations~\eqref{eq:masssimple} and \eqref{eq:momsimple}. The right columns show its reconstruction obtained with the proposed model of Eq.~\eqref{eq:model5}. As such, the left columns display results of the flow solver, whereas the right columns display discrete fields reconstructed from the sum of analytical transient regularized Stokelets contributions.
Qualitative differences between the observed and reconstructed flow disturbances are hardly perceivable, suggesting that the proposed model estimates the particle's self-induced velocity disturbance with good (qualitative) accuracy.\medskip

\begin{figure}\centering
  \includegraphics{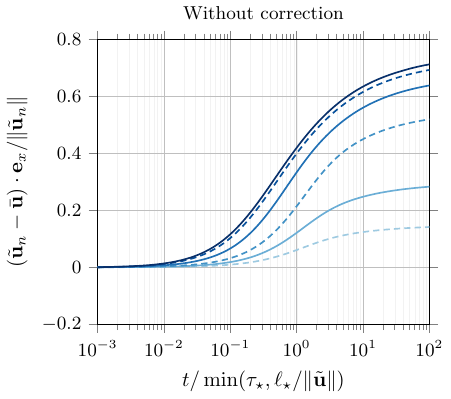}
  \includegraphics{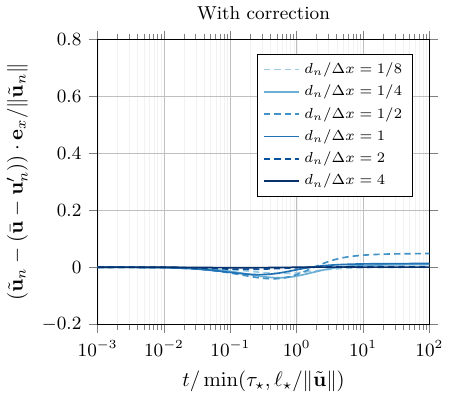}
  \caption{Relative error made in the estimation of the undisturbed velocity at the location of the fixed particle without (left) and with (right) the proposed modeled correction (using Eq.~\eqref{eq:model5}). In this figure, the Reynolds number is fixed to $\mathrm{Re}_n = 0.01$ while $d_n/\Delta x$ is varied in $\smash{\{\frac{1}{8}, \frac{1}{4}, \frac{1}{2}, 1, 2, 4\}}$.}\label{fig:fixedvaryingdx}
\end{figure}

\begin{figure}\centering
  \includegraphics{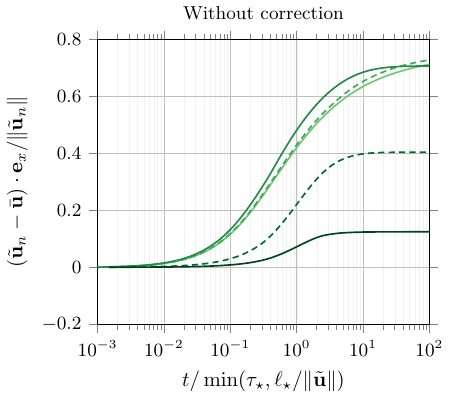}
  \includegraphics{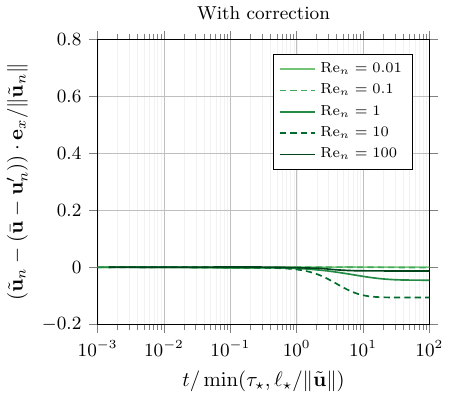}
 \caption{Relative error made in the estimation of the undisturbed velocity at the location of the fixed particle without (left) and with (right) the proposed modeled correction (using Eq.~\eqref{eq:model5}). In this figure, the resolution is fixed to $d_n/\Delta x = 4$ while the Reynolds number is varied in $\smash{\{\frac{1}{100}, \frac{1}{10}, 1, 10, 100\}}$.}\label{fig:fixedvaryingre}
\end{figure}

\begin{figure}\centering
  \includegraphics{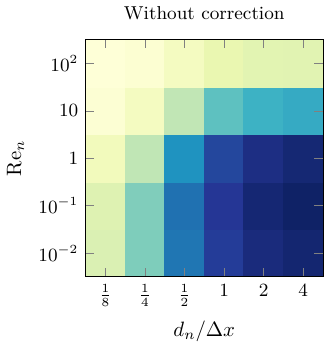}
  \includegraphics{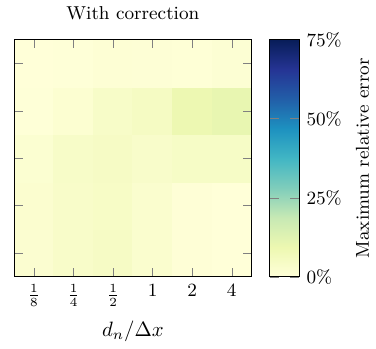}
    \caption{Maximum relative error made in the estimation of the undisturbed velocity at the location of the fixed particle without (left) and with (right) the proposed modeled correction (using Eq.~\eqref{eq:model5}). The values are taken inside the time interval $\left[10^{-3},10^2\right]\times\min(\tau_\star,\ell_\star/\|\tilde{\mathbf{u}}\|)$.}\label{fig:fixedsummary}
\end{figure}

A quantitative analysis of the proposed model is shown in Figures~\ref{fig:fixedvaryingdx}, \ref{fig:fixedvaryingre}, and \ref{fig:fixedsummary}. Figure~\ref{fig:fixedvaryingdx} displays the normalized velocity disturbance interpolated to the center of the particle at $\smash{\mathrm{Re}_n = 0.01}$ and for $\smash{d_n / \Delta x \in \{\frac{1}{8}, \frac{1}{4}, \frac{1}{2}, 1, 2, 4\}}$. In the left plot, the disturbance is calculated as the difference between the $x$-components of the undisturbed velocity, $\smash{\tilde{\vecu}}$, and the filtered velocity, $\smash{\bar{\vecu}}$, interpolated to the particle center. It is then normalized by the norm of the undisturbed velocity $\smash{\tilde{\vecu}}$. In the right plot, this value is corrected by further subtracting the velocity disturbance modeled by Eq.~\eqref{eq:model5}. The left and right plots therefore display the relative errors that would be made in the estimation of the fluid force acting on the particle when not using our model (left) and when using it (right). With infinite time and space resolution and in the limit of vanishing Reynolds number, our model should produce no error, which is illustrated by the limit-case $\smash{(\mathrm{Re}_n = 0.01, d_n / \Delta x = 4)}$ for which the relative error is about $0.2\%$ at most. Figure~\ref{fig:fixedvaryingre} displays the same normalized velocity disturbance interpolated to the center of the particle for $\smash{d_n / \Delta x = 4}$ and at $\smash{\mathrm{Re}_n \in \{\frac{1}{100}, \frac{1}{10}, 1, 10, 100\}}$. Finally, Figure~\ref{fig:fixedsummary} summarizes this study by showing the maximum error made with/without correction across the tested parameter space. Without correction, i.e., by interpolating the filtered velocity at the particle location to estimate drag, a maximum relative error of almost $75\%$ can be made. When using our proposed model to correct this interpolated filtered velocity, the maximum relative error that is made is reduced to about $10\%$ at most.\medskip

Several interesting observations can be made from Figures~\ref{fig:fixedvaryingdx}, \ref{fig:fixedvaryingre}, and \ref{fig:fixedsummary}. First, it is clear that when no correction is applied, the magnitude of the particle's self-induced velocity disturbance increases with the ratio $\smash{d_n / \Delta x}$ and decreases with increasing Reynolds number. This is a well-known result of the VF-EL literature \cite{Ireland2017,Horwitz2018,Balachandar2019,Evrard2020a,Evrard2021}. 
Second, it is apparent that at high mesh resolution, our model's bottom performance is reached for $\smash{\mathrm{Re}_n}$ around 10. As shown by the evolution of the coefficient $\Psi_\mathscr{W}$ as a function of the Reynolds number (see Eq.~\eqref{eq:psioseen}), the magnitude of the particle's self-induced velocity disturbance decreases proportional to the inverse of the Reynolds number. The case $\smash{\mathrm{Re}_n \sim 10}$ thus corresponds to the ``worst-case'' scenario in which the magnitude of the self-induced velocity disturbance is still relatively significant while non-linear effect are also important, therefore approaching the limits of our model built upon linearized governing equations.

\subsection{Oscillating particle in uniform flow}\label{sec:oscillpart}
We now consider the case of a particle subject to a prescribed oscillatory motion in a uniform flow with the constant flud velocity $\smash{\tilde{\vecu}}$. This case aims to test the performance of our model for predicting the particle's self-induced velocity disturbance in a highly transient environment where the resultant of the fluid forces acting on the particle is not aligned with the undisturbed flow velocity.\medskip

The numerical setup and considered parameter space are identical to those of Section~\ref{sec:fixedpart}. The main difference lies in the particle having a prescribed motion instead of being fixed. This prescribed motion is given as
\begin{equation}
  \vecX_{n}(t) = 5 d_n
  \begin{bmatrix}
    \sin(4\omega t) \\
    \sin(4\omega t)\cos(\omega t) \\
    \sin(4\omega t)\sin(\omega t)
  \end{bmatrix} \, ,
\end{equation}
with 
\begin{equation}
  \omega = \dfrac{\pi \|\tilde{\vecu}\|}{25 d_n} \, .
\end{equation}
This corresponds to a period of oscillation, $\smash{T = \pi / 2\omega}$, approximately varying between $\smash{0.02 \tau_\nu}$ and $\smash{200 \tau_\nu}$ across the chosen parameter space.
The velocity of the particle is given as
\begin{equation}
  \vecU_{n}(t) = 5 d_n 
  \begin{bmatrix}
    4 \omega \cos(4\omega t)\\
    4 \omega \cos(4\omega t)\cos(\omega t) - \omega \sin(4\omega t) \sin(\omega t)\\
    4 \omega \cos(4\omega t)\sin(\omega t) + \omega \sin(4\omega t) \cos(\omega t)
  \end{bmatrix} \, .
\end{equation}
The particle feeds back to the fluid a momentum contribution corresponding to the opposite of the steady drag force acting on the particle,
\begin{equation}
  \vecf(t) = -3\pi \mu d_n \left(\tilde{\vecu} - \vecU_{n}(t)\right) f(\mathrm{Re}_n) \mathscr{W}(\|\vecx - \vecX_{n}(t)\|) \, ,
\end{equation}
where $f(\mathrm{Re}_n)$ is the \citet{Schiller1933} empirical correction factor given in Eq.~\eqref{eq:schillernaumann}. The solver timestep is chosen as the minimum between
\begin{equation}
  \Delta t = \dfrac{d_n}{10\|\tilde{\vecu}\|} \, ,
\end{equation}
and the timestep enforcing $\text{CFL} = 0.1$.\medskip

\begin{figure} \centering
	\includegraphics{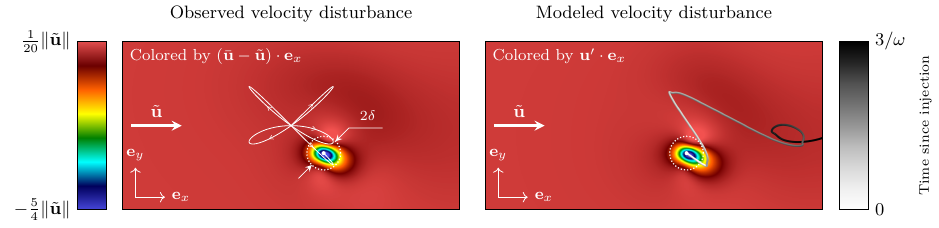}\\
  \includegraphics{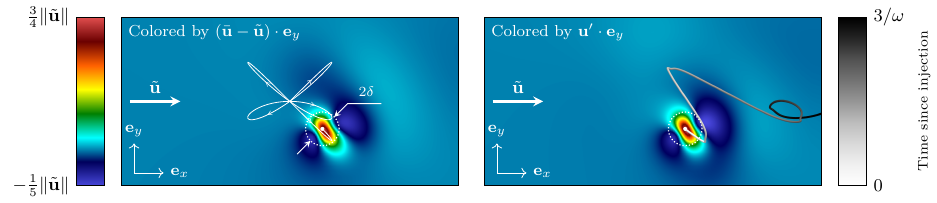}\\
  \includegraphics{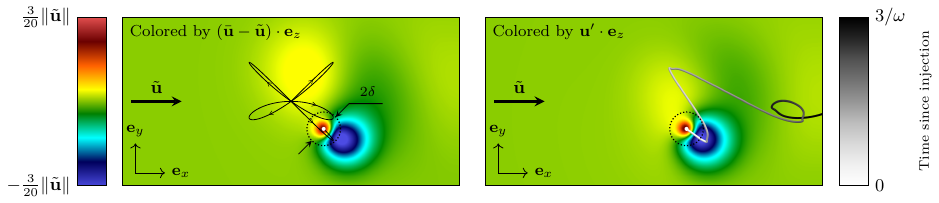}
  \caption{Snapshots of the volume-filtered simulation of a particle oscillating with a prescribed trajectory in a uniform flow with constant far-field velocity $\smash{\tilde{\vecu}}$ at the Reynolds number $\smash{\mathrm{Re}_n = d_n \|\tilde{\vecu}\| / \nu =  1}$. The Wendland filter $\smash{\smash{\mathscr{W}}}$ defined in Eq.~\eqref{eq:defwendland} is used with $\delta = 2d_n$ and with the resolution $\smash{d_n / \Delta x = 4}$. These snapshots are taken at a simulation time ${t \simeq 11/3\omega}$. The left column shows the particle's self-induced velocity disturbance obtained by subtracting the undisturbed flow velocity $\smash{\tilde{\vecu}}$ from the filtered velocity $\smash{\bar{\vecu}}$ solution to the volume-filtered governing equations. The right column shows the reconstruction of the particle's self-induced velocity disturbance with the model proposed in Eq.~\eqref{eq:model5}. It also displays the stream of ``Stokeslet sources'' advected by the background flow and used by the model, which are colored in a gray scale according to the difference between the current time and their time of injection. The rows correspond, from top to bottom, to the $x$-, $y$-, and $z$-components of the velocity disturbance.} \label{fig:oscillatingre1}
\end{figure}

\begin{figure} \centering
  \begin{tabular}{r}
    \includegraphics{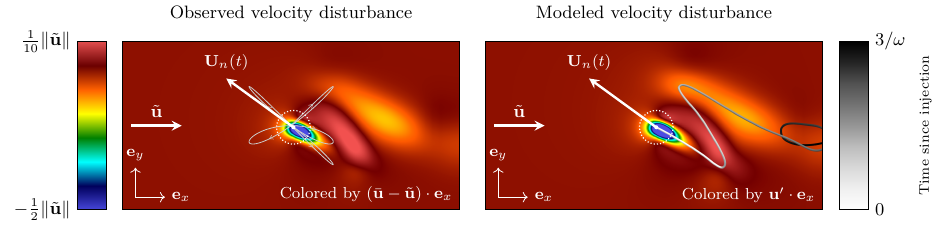}\\
    \includegraphics{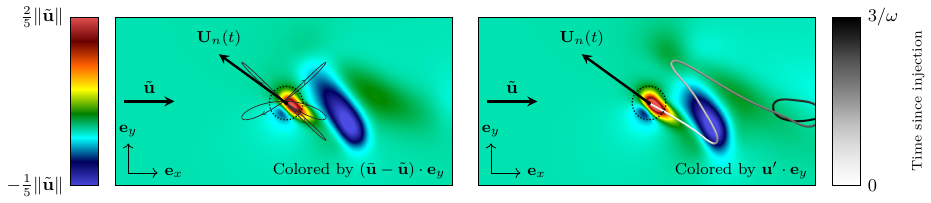}\\
    \includegraphics{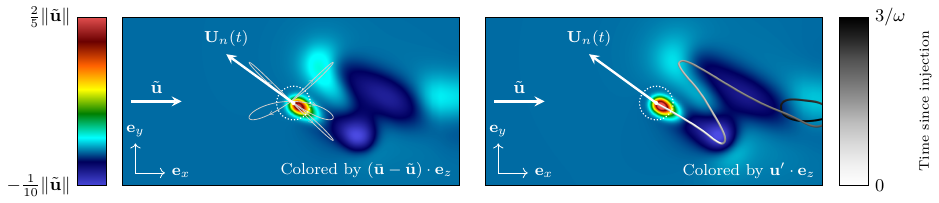}
  \end{tabular}
  \caption{Snapshots of the volume-filtered simulation of a particle oscillating with a prescribed trajectory in a uniform flow with constant far-field velocity $\smash{\tilde{\vecu}}$ at the Reynolds number $\smash{\mathrm{Re}_n = d_n \|\tilde{\vecu}\| / \nu =  10}$. The Wendland filter $\smash{\smash{\mathscr{W}}}$ defined in Eq.~\eqref{eq:defwendland} is used with $\delta = 2d_n$ and with the resolution $\smash{d_n / \Delta x = 4}$. These snapshots are taken at a simulation time ${t \simeq 4/\omega}$. The left column shows the particle's self-induced velocity disturbance obtained by subtracting the undisturbed flow velocity $\smash{\tilde{\vecu}}$ from the filtered velocity $\smash{\bar{\vecu}}$ solution to the volume-filtered governing equations. The right column shows the reconstruction of the particle's self-induced velocity disturbance with the model proposed in Eq.~\eqref{eq:model5}. It also displays the stream of ``Stokeslet sources'' advected by the background flow and used by the model, which are colored in a gray scale according to the difference between the current time and their time of injection. The rows correspond, from top to bottom, to the $x$-, $y$-, and $z$-components of the velocity disturbance.} \label{fig:oscillatingre10}
\end{figure}

Figures~\ref{fig:oscillatingre1} and \ref{fig:oscillatingre10} show snapshots of the simulations at $\smash{\mathrm{Re}_n = 1}$ and $\smash{\mathrm{Re}_n = 10}$, with $d_n / \Delta x = 4$. The left columns show the components of the particle's self-induced velocity disturbance obtained by subtracting the undisturbed velocity from the filtered velocity solution to Eqs.~\eqref{eq:masssimple} and \eqref{eq:momsimple}. They also show the projection of the particle's trajectory in the $(x,y)$ plane up until the time at which the snapshot is taken. The right columns show the particle's self-induced velocity disturbance reconstructed with the proposed model of Eq.~\eqref{eq:model5}. They also show the stream of transient Stokeslet sources that are advected by the background flow, colored by their time of injection relative to the current snapshot time.
Qualitative differences between the velocity disturbance observed in the simulation and that estimated with the proposed model are hardly perceivable, indicating that an accurate recovery of the undisturbed velocity from the filtered flow is possible even in this highly transient case with significant inertial effects.\medskip

\begin{figure}\centering
  \includegraphics{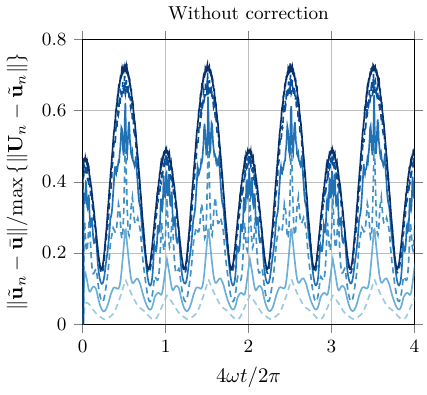}
  \includegraphics{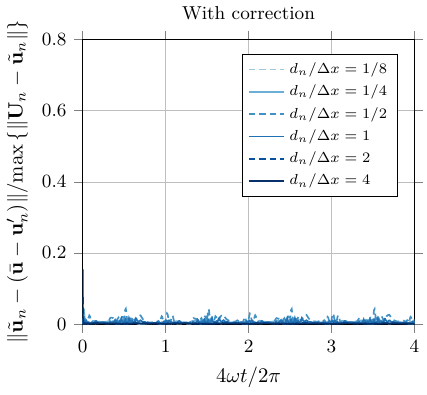}
  \caption{Error made in the estimation of the undisturbed velocity at the location of the oscillating particle without (left) and with (right) the proposed modeled correction (using Eq.~\eqref{eq:model5}), normalized by the maximum relative velocity between the particle and undisturbed flow. In this figure, the Reynolds number is fixed to $\mathrm{Re}_n = 0.01$ while $d_n/\Delta x$ is varied in $\smash{\{\frac{1}{8}, \frac{1}{4}, \frac{1}{2}, 1, 2, 4\}}$. Four periods of oscillation of the particle are considered.}\label{fig:oscillatingvaryingdx}
\end{figure}

\begin{figure}\centering
  \includegraphics{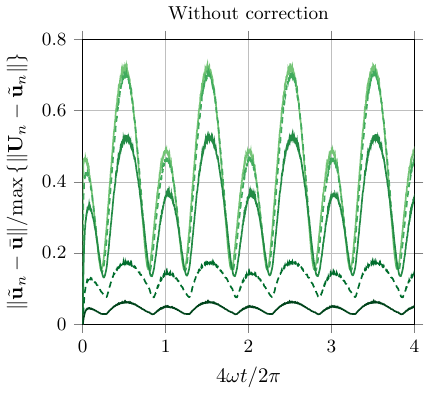}
  \includegraphics{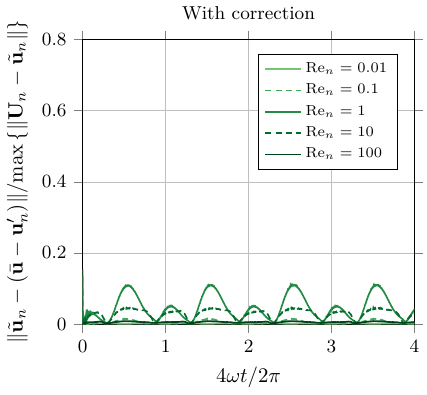}
\caption{Error made in the estimation of the undisturbed velocity at the location of the oscillating particle without (left) and with (right) the proposed modeled correction (using Eq.~\eqref{eq:model5}), normalized by the maximum relative velocity between the particle and undisturbed flow. In this figure, the resolution is fixed to $d_n/\Delta x = 4$ while the Reynolds number is varied in $\smash{\{\frac{1}{100}, \frac{1}{10}, 1, 10, 100\}}$. Four periods of oscillation of the particle are considered.}\label{fig:oscillatingvaryingre}
\end{figure}

\begin{figure}\centering
  \includegraphics{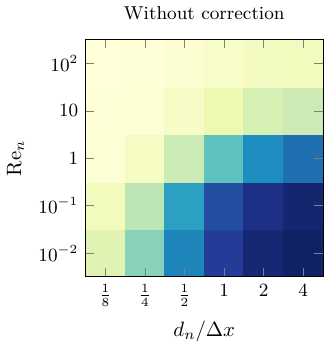}
  \includegraphics{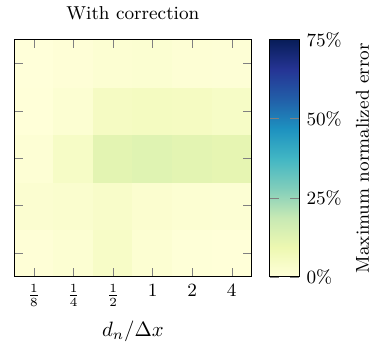}
    \caption{Maximum error made in the estimation of the undisturbed velocity at the location of the fixed particle without (left) and with (right) the proposed modeled correction (using Eq.~\eqref{eq:model5}), normalized by the maximum relative velocity between the particle and undisturbed flow. The values are taken inside one period of oscillation of the particle.}\label{fig:oscillatingsummary}
\end{figure}

A quantitative analysis of the proposed model is shown in Figures~\ref{fig:oscillatingvaryingdx}, \ref{fig:oscillatingvaryingre}, and \ref{fig:oscillatingsummary}. Similarly to the previous case of a fixed particle, Figure~\ref{fig:oscillatingvaryingdx} displays the normalized velocity disturbance interpolated to the center of the particle at $\smash{\mathrm{Re}_n = 0.01}$ and for $\smash{d_n / \Delta x \in \{\frac{1}{8}, \frac{1}{4}, \frac{1}{2}, 1, 2, 4\}}$. In the left column is plotted the norm of the difference between the undisturbed velocity, $\smash{\tilde{\vecu}}$, and the filtered velocity, $\smash{\bar{\vecu}}$, interpolated to the particle center. It is then normalized by the maximum magnitude of the relative velocity between the particle and the undisturbed flow. In the right column, this value is corrected by further subtracting the velocity disturbance modeled by Eq.~\eqref{eq:model5}. The left and right plots therefore display the relative errors that would be made in the estimation of the fluid force acting on the particle when not using our model (left) and when using it (right). Figure~\ref{fig:oscillatingvaryingre} displays the same normalized velocity disturbance interpolated to the center of the particle for $\smash{d_n / \Delta x = 4}$ and at $\smash{\mathrm{Re}_n \in \{\frac{1}{100}, \frac{1}{10}, 1, 10, 100\}}$. Finally, Figure~\ref{fig:oscillatingsummary} summarizes this study by showing the maximum error made with/without correction across the tested parameter space. Without correction, i.e., by interpolating the filtered velocity at the particle location to estimate drag, a maximum normalized error of about $75\%$ can be made, similarly to the case of a fixed particle. When using our proposed model to correct this interpolated filtered velocity, the maximum normalized error that is made is reduced to about $13\%$. This demonstrates that the proposed model is not only able to accurately reconstruct the particle's self-induced velocity disturbance for a particle with constant velocity, but also in highly transient and inertia dominated environments. \medskip

\subsection{Particle settling under the influence of gravity}
We now consider the case of an isolated particle settling under the influence of gravity in a very large domain filled with quiescent fluid (i.e., the undisturbed velocity for this particle is $\smash{\tilde{\vecu} = \mathbf{0}}$). We only consider the steady drag force and the gravitational force acting on the particle. In such a case, Newton's second law, Eq.~\eqref{eq:Ppos}, becomes
\begin{equation} 
  \dfrac{\mathrm{d}\vecU_n(t)}{\mathrm{d}t} = -\dfrac{\vecU_n(t)}{\tau_n} + \mathbf{g} \, ,
\end{equation}
where $\tau_n$ is the particle time-scale given as
\begin{equation}
  \tau_n = \dfrac{\rho_n d_n^2}{18\mu f(\mathrm{Re}_n)} \, ,
\end{equation}
and with $\smash{\mathrm{Re}_n}$ the Reynolds number of the particle defined as
\begin{equation}
  \mathrm{Re}_n = \dfrac{d_n \|\vecU_n(t)\|}{\nu} \, .  \label{eq:reynoldsnumber}
\end{equation}
Note that $f(\mathrm{Re}_n)$ is again the \citet{Schiller1933} empirical correction factor given in Eq.~\eqref{eq:schillernaumann}. The terminal velocity reached by the particle is given by
\begin{equation}
  \vecU_{n,\infty} = \lim\limits_{t\to\infty}\vecU_n(t) = \tau_n \mathbf{g} \, .
\end{equation}
This case has been studied in the vast majority of papers that propose models for recovering the undisturbed velocity in volume-averaged Euler-Lagrange simulations \cite[e.g.,][]{Ireland2017,Balachandar2019,Evrard2020a}. In order to quantify the influence of transient effects in the development of the particle's self-induced flow disturbance, we introduce a Stokes number based on the viscous time-scale of the regularization/filtering kernel, $\tau_\nu$, defined in Eq.~\eqref{eq:defviscoustau}. This Stokes number reads as
\begin{equation}
  \mathrm{St}_n = \dfrac{\tau_n}{\tau_\nu} = \dfrac{\rho_n d_n^2}{18\rho \ell^2 f(\mathrm{Re}_n)} \, . \label{eq:stokesnumber}
\end{equation}
When $\smash{\mathrm{St}_n} \ll 1$, the rate at which the particle's self-induced flow disturbance develops is slower than the rate at which the particle relaxes to its terminal velocity. Therefore, one should expect that it is critical for a model recovering the undisturbed velocity to account for transient effects, so as to accurately predict the time evolution of the particle. When $\smash{\mathrm{St}_n} \gg 1$, on the other hand, one should expect steady models as proposed in \cite{Balachandar2019,Evrard2020a} to be sufficient for accurately predicting the evolution of the particle.\medskip

\begin{figure}\centering
  \includegraphics{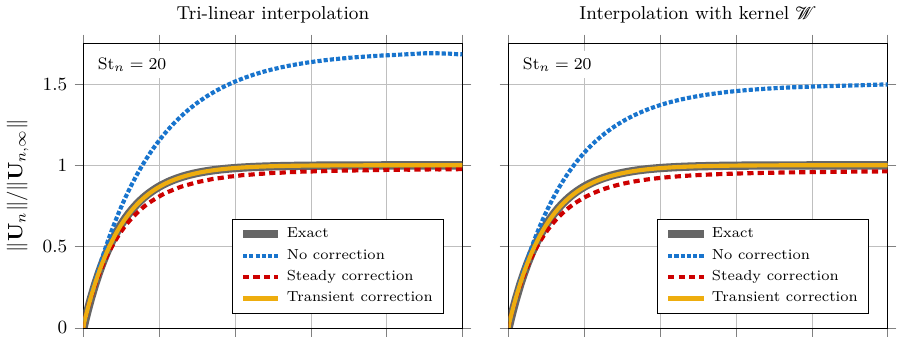}
  \includegraphics{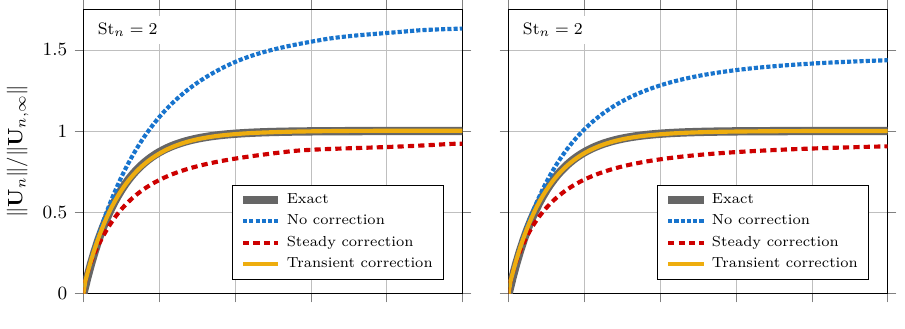}
  \includegraphics{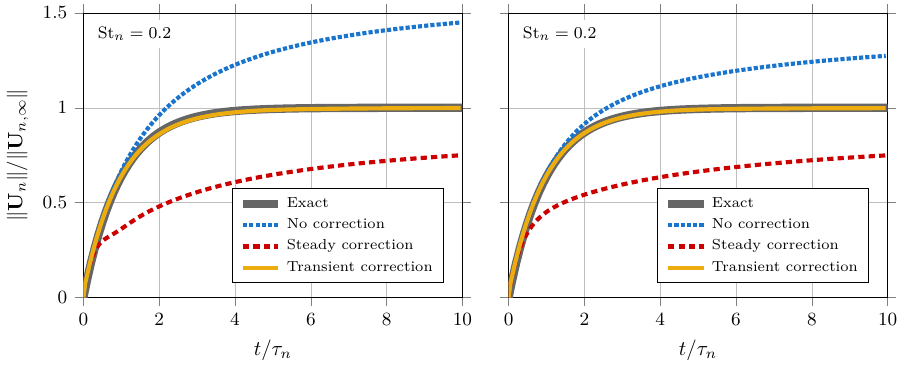}
  \caption{Velocity of a particle settling under the influence of gravity in the VF-EL framework, normalized by its terminal velocity. The Reynolds number of the particle, based on its terminal velocity, is $\smash{\mathrm{Re}_n = 0.1}$. From top to bottom, the Stokes number of the particle, based on the viscous time-scale of the filtering kernel, is $\smash{\mathrm{St}_n \in \{20,2,0.2\}}$. In the left column, velocity is interpolated at the particle location using tri-linear interpolation. In the right column, the filtering kernel is used as interpolation kernel too. Each plot reports the exact expected particle velocity, the particle velocity when no correction is applied to the interpolated filtered velocity, the particle velocity when the steady corrections of \citet{Evrard2021} and \citet{Evrard2020a} are applied to the interpolated filtered velocity, and the particle velocity when the model of Eq.\eqref{eq:model5} is used to correct the filtered velocity.} \label{fig:settlingStvarying}
\end{figure}

Figure~\ref{fig:settlingStvarying} displays the predicted settling velocity of the particle normalized by its terminal velocity. The Wendland kernel $\smash{\mathscr{W}}$ defined in Eq.~\eqref{eq:defwendland} is used to regularize the momentum exchange term $\vecf$ in Eqs.~\eqref{eq:masssimple} and \eqref{eq:momsimple}. The radius of the kernel's support is chosen equal to two particle diameters, $\delta = 2d_n$, and the resolution $d_n/\Delta x = 2$ is employed in the vicinity of the particle. This corresponds to $8$ cells across the filter's support. The flow solver timestep is chosen as $\smash{\Delta t = \min(\tau_\nu/2,\tau_n/20,\Delta x/2\|\vecU_{n,\infty}\|)}$. The Reynolds number of the particle based on its terminal velocity is set to $\smash{\mathrm{Re}_n = 0.1}$. Each row of Figure~\ref{fig:settlingStvarying} corresponds to a Stokes number $\smash{\mathrm{St}_n \in \{0.2, 2, 20\}}$. Each column of Figure~\ref{fig:settlingStvarying} corresponds to a choice of operator for interpolating the fluid velocity at the location of the particle. In the left column, tri-linear interpolation is used. In the right column, the regularization kernel $\smash{\mathscr{W}}$ is used as interpolation kernel, as proposed in~\cite{Evrard2020a}. In each plot of Figure~\ref{fig:settlingStvarying}, we report four results:
\begin{enumerate}
  \item The ``exact'' velocity of the settling particle, predicted using the exact undisturbed velocity $\smash{\tilde{\vecu} = \mathbf{0}}$.
  \item The velocity of the settling particle obtained when no correction is employed, i.e., the filtered fluid velocity $\smash{\bar{\vecu}}$, solution to Eqs.~\eqref{eq:masssimple} and \eqref{eq:momsimple}, is used to approximate the undisturbed velocity $\smash{\tilde{\vecu}}$.
  \item The velocity of the settling particle obtained when a steady correction is employed to approximate the undisturbed velocity $\smash{\tilde{\vecu}}$: With tri-linear interpolation, the steady correction of \citet{Evrard2021} is used; With the interpolation kernel $\smash{\mathscr{W}}$, the steady correction of \citet{Evrard2020a} is used.
  \item The velocity of the settling particle obtained when the model of Eq.~\eqref{eq:model5} is used to approximate the particle's self-induced velocity disturbance for recovering the undisturbed velocity $\smash{\tilde{\vecu}}$.
\end{enumerate}

From Figure~\ref{fig:settlingStvarying}, one can observe that the accuracy of a steady correction model such as proposed in \cite{Balachandar2019,Evrard2020a,Evrard2021} improves with increasing Stokes number (as defined per Eq.~\eqref{eq:stokesnumber}). As expected, as well, applying no correction to the interpolated filtered fluid velocity leads to a large overestimation of the particle's terminal velocity. Finally, for all Stokes numbers considered, the proposed transient model works as intended and approximates the transient evolution of the particle's self induced velocity disturbance with high accuracy. This leads to a near-exact prediction of the particle's terminal velocity, but also of the evolution of its velocity over time.\medskip

To provide some illustrative context, a sand particle of diameter $d_n = 0.1\ \mathrm{mm}$ sedimenting in water would display, at terminal velocity, a Reynolds number $\smash{\mathrm{Re}_n = \mathcal{O}(1)}$ and a Stokes number $\smash{\mathrm{St}_n = \mathcal{O}(0.1)}$ according to the definitions of Eqs.~\eqref{eq:reynoldsnumber} and \eqref{eq:stokesnumber}. These conditions are close to those of the bottom row of Figure~\ref{fig:settlingStvarying}, in which no correction led to a significant overestimation of the particle's terminal velocity, while a steady correction led to its significant underestimation. Hence, the use of a transient model such as the one proposed in Eq.~\eqref{eq:model5} for approximating the particle's self-induced velocity disturbance may radically improve the accuracy of the simulation of particle-laden flows with such physical parameters.

\subsection{Truncation of the forcing history}\label{sec:history}
We have shown in Section~\ref{sec:timelimit} that the contribution of the $m^\mathrm{th}$ previous forcing instance in the discrete time-convolution of Eq.~\eqref{eq:model5} decays proportional to $\smash{m^{-2/3}}$. This means that the sum in Eq.~\eqref{eq:model5} may be truncated without significantly affecting the accuracy of the model. We provide in Figure~\ref{fig:ratiotophat} an estimation of the relative importance of the $m^\mathrm{th}$ previous forcing instance relative to the most recent one, for a regularization with the Gaussian kernel and for a range of values of the ratio $\smash{\Delta t / \tau_\nu}$.\medskip

In this section, we consider the previously described case of settling particle with $\smash{\mathrm{Re}_n = 0.1}$ and $\smash{\mathrm{St}_n = 20}$, and gradually decrease the threshold beyond which previous forcing instances are discarded in the estimation the particle's self-induced velocity disturbance. To be more precise, the terms of Eq.~\eqref{eq:model5} for which $t - \tk > \Upsilon$ are dropped in the estimation of $\smash{\vecu^\prime}$. We consider the limits $\smash{\Upsilon \in \{\infty, 100, 50, 25, 12.5, 6.25\} \times \tau_\nu}$. Note that for this specific case, the particle time-scale is $20$ times the filter's viscous time-scale, $\smash{\tau_n = 20 \tau_\nu}$.\medskip

\begin{figure}\centering
  \includegraphics{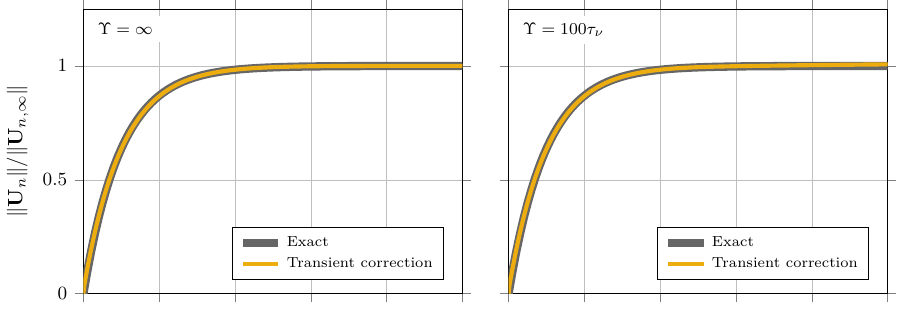}
  \includegraphics{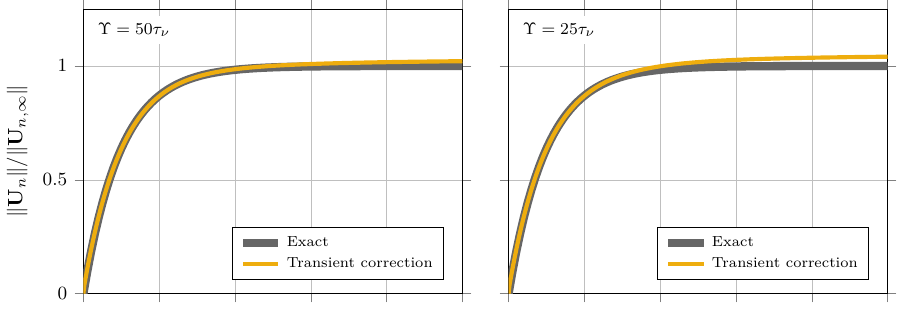}
  \includegraphics{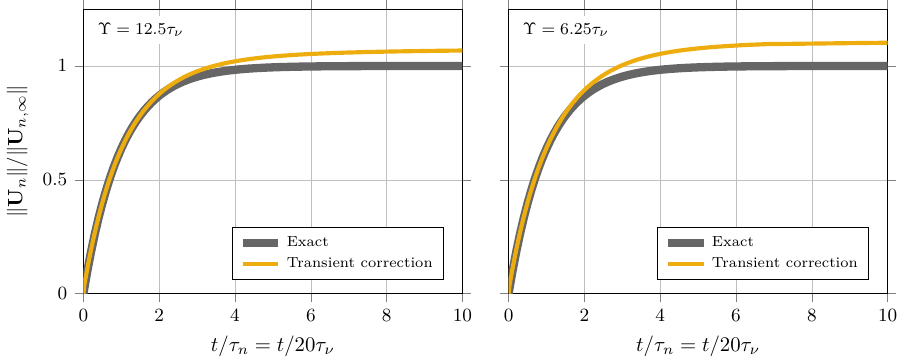}
  \caption{Velocity of a particle settling under the influence of gravity in the VF-EL framework, normalized by its terminal velocity. The Reynolds number of the particle, based on its terminal velocity, is $\smash{\mathrm{Re}_n = 0.1}$ and its Stokes number of the particle, based on the viscous time-scale of the filtering kernel, is $\smash{\mathrm{St}_n = 20}$. Velocity is interpolated to the particle location using tri-linear interpolation. From top left to bottom right, the threshold beyond which previous forcing instances are discarded in the estimation the velocity disturbance is gradually decreased, from $\Upsilon = \infty$ to $\Upsilon = 6.25 \tau_\nu$. The particle velocity obtained using the (truncated) model of Eq.~\eqref{eq:model5} to correct the filtered velocity is compared against the exact expected particle velocity.} \label{fig:settlingtruncation}
\end{figure}

Figure~\ref{fig:settlingtruncation}, compares the particle velocity predicted using the truncated transient model correction to the exact particle velocity. As expected, reducing $\Upsilon$ -- which amounts to reducing the amount of terms considered in the sum of Eq.~\eqref{eq:model5} -- leads to an error in the estimation of the particle's self-induced velocity disturbance. In the extreme case we have considered, for which $\smash{\Upsilon = 6.25 \tau_\nu}$, only $12$ previous forcing instances are stored for estimating the particle's self-induced velocity disturbance with Eq.~\eqref{eq:model5}. In this case, the particle's terminal velocity is predicted with an error of about $10\%$ while no correction at all leads to an error of about $75\%$.\medskip

Overall, the choice of truncating the forcing history of a given particle for estimating its self-induced velocity disturbance is one that balances the level of accuracy one wishes to reach with the computational costs one can afford. Truncating the sum in Eq.~\eqref{eq:model5} will reduce the computational cost and memory footprint of the model, but may results in undisturbed velocity predictions that are significantly erroneous. A transient correction model that considers as few as one or two previous forcing instances is better than no correction at all, and is relatively cheap. However, for certain flow configurations, it may not perform as well as the steady correction models previously proposed in the literature \cite[e.g.,][]{Balachandar2019,Evrard2020a}, which bear computational costs of a similar order of magnitude as one or two transient correction contributions.

\section{Conclusions}
\label{sec:conclusions}

This manuscript proposes a new model for recovering the undisturbed velocity associated with a particle in volume-filtered Euler-Lagrange simulations of particle-laden flows. This model is built upon the linearization of the equations governing the flow perturbation induced by a particle, whose solution can then be written in the form of a temporal and spatial convolution integral involving known analytical operators. These convolution integrals are discretized with first-order accuracy in time and second-order accuracy in space. The resulting approximate solution of the particle's self-induced velocity disturbance is a linear combination of regularized transient Stokeslet contributions, and does not rely on any ad hoc or empirical parameter. The model accounts for the transient development of the velocity disturbance, is shown to provide accurate estimations at finite particle Reynolds numbers, and does not require the momentum feedback force and the particle's relative velocity vector to be aligned, enabling the consideration of fluid forces other than the steady drag force.\medskip

The proposed model is first tested on VF-EL cases of particles with prescribed motion. These tests consider particle Reynolds numbers spanning four orders of magnitude, and mesh resolutions ranging from 1 cell (PSI-CELL equivalent) to 16 cells across the filter support. Over these 60 test-cases, some of which display significant transient and inertial effects in the development of the particle's self-induced flow disturbance, the proposed model consistently enables the recovery of the undisturbed velocity with a relative error on the order of a few percents. A maximum relative error of about $13\%$ is obtained with the proposed model, whereas an error of up to $75\%$ can be reached without any correction. The model is then tested on cases of particles settling under the influence of gravity, varying the Stokes number defined with respect to the viscous time-scale of the filter. The proposed model then consistently outperforms a steady correction model, and enables high-accuracy prediction of the particle's velocity.\medskip

The high degree of accuracy of the proposed model comes at the cost of seeding and keeping track of fluid tracers along the trajectory of each particle. In order to keep the associated computational cost low, the discrete time convolution sum of Eq.~\eqref{eq:model5} can be truncated. The choice of a threshold beyond which such truncation is applied is one that must balance available computational resources with the desired accuracy. It is also one that depends on the specificities of the case under consideration, especially on the ratio of the solver timestep over the viscous time-scale of the filter.

\newpage
\appendix
\section{Closed-form expressions of some regularized transient Stokeslet operators}
\label{apdx:manualconv}
\subsection{Top-hat filter}\label{apdx:manualconvtophat}
Consider the top-hat radial filter kernel
\begin{equation}
	\mathscr{H}(r) = \dfrac{3}{4\pi \delta^3}\left\{ \begin{array}{ll} 1 & r < \delta \\ 0 & r \ge \delta\end{array}\right. \, . \label{eq:kerneltophat}
\end{equation}
At $\vecx = \boldsymbol{0}$, the convolution of the transient persistent Stokeslet tensor operator with $\mathscr{H}$ yields the diagonal operator
\begin{equation}
	\integral{\vecG}_\mathscr{H} (\boldsymbol{0},t) = \identity\,\mathscr{S}_{\mathscr{H}_0} (t) \, ,
\end{equation}
where $\smash{\mathscr{S}_{\mathscr{H}_0}}$ is defined as in Eq.~\eqref{eq:intoriginstokeslet},
\begin{equation}
	\mathscr{S}_{\mathscr{H}_0} (t) = \frac{4\pi}{\mu} \dfrac{3}{4\pi \delta^3} \int_0^\delta \left(r^2\integral{\Hh}_1(r,t) + \frac{r^4}{3}\integral{\Hh}_2(r,t)\right)  \, \mathrm{d}r \, . \label{eq:intoriginstokeslettophat}
\end{equation}
This reads as
\begin{equation}
	\mathscr{S}_{\mathscr{H}_0} (t) = \dfrac{1}{4\pi \delta \mu} \left( 1 - \left(1-\frac{2\nu t}{\delta^2}\right) \erf \left( \frac{\delta}{\sqrt{4\nu t}}\right) - \frac{2}{\delta}\sqrt{\frac{\nu t}{\pi}} \exp\left( -\frac{\delta^2}{4\nu t} \right)\right) \, .
\end{equation}
The corresponding quantity for the steady regularized Stokeslet is given as
\begin{equation}
	\lim\limits_{t\to\infty}\mathscr{S}_{\mathscr{H}_0} (t) = \dfrac{1}{4\pi \delta \mu} \, .
\end{equation}
Note that this steady solution corresponds to the solution derived by \citet{Evrard2021} in the limit of vanishing Reynolds number.

\subsection{Gaussian filter}
Consider the gaussian filter kernel
\begin{equation}
	\mathscr{G}(r) = \dfrac{1}{(2\pi\sigma^2)^{3/2}} \exp\left( \frac{-r^2}{2\sigma^2}\right) \, . \label{eq:kernelgaussian}
\end{equation}
At $\vecx = \boldsymbol{0}$, the convolution of the transient persistent Stokeslet tensor operator with $\mathscr{G}$ yields the diagonal operator
\begin{equation}
	\integral{\vecG}_\mathscr{G} (\boldsymbol{0},t) = \identity\,\mathscr{S}_{\mathscr{G}_0} (t) \, ,
\end{equation}
where $\smash{\mathscr{S}_{\mathscr{G}_0}}$ is defined as in Eq.~\eqref{eq:intoriginstokeslet},
\begin{equation}
	\mathscr{S}_{\mathscr{G}_0} (t) = \frac{4\pi}{\mu} \dfrac{1}{(2\pi\sigma^2)^{3/2}} \int_0^\infty \exp\left( \frac{-r^2}{2\sigma^2}\right)\left(r^2\integral{\Hh}_1(r,t) + \frac{r^4}{3}\integral{\Hh}_2(r,t)\right)  \, \mathrm{d}r \, . \label{eq:intoriginstokesletgaussian}
\end{equation}
This reads as
\begin{equation}
	\mathscr{S}_{\mathscr{G}_0} (t) = \frac{1}{3\pi\sqrt{2\pi}\sigma\mu} \left(1 - \frac{\sigma}{\sqrt{2\nu t + \sigma^2}} \right)\, .
\end{equation}
The corresponding quantity for the steady regularized Stokeslet is given as
\begin{equation}
	\lim\limits_{t\to\infty}\mathscr{S}_{\mathscr{G}_0} (t) = \frac{1}{3\pi\sqrt{2\pi}\sigma\mu} \, .
\end{equation}
Note that this steady solution was derived by \citet{Balachandar2019}.

\subsection{Wendland filter}\label{appendix:wendland}
Consider the compactly supported, polynomial Wendland filter kernel \citep{Wendland1995}
\begin{equation}
	\mathscr{W}(r) = \dfrac{21}{2\pi \delta^3} \left\{ \begin{array}{ll} \left( \dfrac{4r}{\delta} + 1 \right)\left( 1-\dfrac{r}{\delta} \right)^4 & r < \delta \\ 0 & r \ge \delta\end{array}\right. \, . \label{eq:defwendland}
\end{equation}
At $\vecx = \boldsymbol{0}$, the convolution of the transient persistent Stokeslet tensor operator with $\mathscr{W}$ yields the diagonal operator
\begin{equation}
	\integral{\vecG}_\mathscr{W} (\boldsymbol{0},t) = \identity\,\mathscr{S}_\mathscr{W} (t) \, ,
\end{equation}
where $\smash{\mathscr{S}_{\mathscr{W}_0}}$ is defined as in Eq.~\eqref{eq:intoriginstokeslet},
\begin{equation}
	\mathscr{S}_{\mathscr{W}_0} (t) = \frac{4\pi}{\mu} \dfrac{21}{2\pi \delta^3} \int_0^\delta \left( \dfrac{4r}{\delta} + 1 \right)\left( 1-\dfrac{r}{\delta} \right)^4\left(r^2\integral{\Hh}_1(r,t) + \frac{r^4}{3}\integral{\Hh}_2(r,t)\right)  \, \mathrm{d}r \, . \label{eq:intoriginstokesletwendland}
\end{equation}
This reads as
\begin{equation}\begin{split}
	\mathscr{S}_{\mathscr{W}_0} (t) = \frac{1}{2\pi \delta \mu} & \left(1 + \sqrt{\frac{\nu t}{\pi}} \left( 3584 \delta^{-5} \nu^2 t^2 + 6144 \delta^{-7} \nu^3 t ^3\right) \right.\\
	 & \quad \quad \left. - \left(1 - 14\delta^{-2} \nu t + 420 \delta^{-4}\nu^2 t^2 +4200 \delta^{-6}\nu^3 t^3\right) \erf \left( \frac{\delta}{\sqrt{4\nu t}}\right) \right. \\
	 & \quad \quad \quad \left. - \frac{2}{\delta}\sqrt{\frac{\nu t}{\pi}} \left(1 - 16\delta^{-2} \nu t + 460 \delta^{-4} \nu^2 t^2 + 3072 \delta^{-6}\nu^3 t^3 \right)\exp\left( -\frac{\delta^2}{4\nu t} \right)
	 \right)\, .
	 \end{split}
\end{equation}
The corresponding quantity for the steady regularized Stokeslet is given as
\begin{equation}
	\lim\limits_{t\to\infty}\mathscr{S}_{\mathscr{W}_0} (t) = \frac{1}{2\pi \delta \mu} \, . \label{eq:limwendlandstokeslet}
\end{equation}
Note that this steady solution was derived by \citet{Evrard2020a}.

\section*{Acknowledgements}
This work has been funded by the Deutsche Forschungsgemeinschaft (DFG), grant number 457515061. The authors are grateful to Olivier Desjardins for providing guidance and resources that have supported the work presented in this manuscript. The authors are also grateful to the National Center for Supercomputing Applications (NCSA) at the University of Illinois Urbana-Champaign for providing access to HPC resources for the current work. Finally, the authors thank Sivaramakrishnan Balachandar and Jesse Capecelatro for enlightening and constructive discussions on the topic of undisturbed velocity recovery over the years.

\bibliographystyle{model1-num-names}
\bibliography{ms.bib}

\end{document}